\documentclass[fleqn,usenatbib]{mnras}

\PassOptionsToPackage{nonamebreak}{natbib}   

\usepackage{newtxtext,newtxmath}

\usepackage[T1]{fontenc}
\usepackage{ae,aecompl}

\usepackage[normalem]{ulem}

\usepackage{etoolbox}
\makeatletter
\patchcmd\@combinedblfloats{\box\@outputbox}{\unvbox\@outputbox}{}{%
  \errmessage{\noexpand\@combinedblfloats could not be patched}%
}%
\makeatother

\newcommand\msun{M_{\odot}}
\newcommand\Zsun{Z_{\odot}}

\usepackage{graphicx}	
\usepackage{amsmath}	
\usepackage{amssymb}	

\newcommand\A[1]{{color{black} #1 }}

\title[BBH in the Milky Way]{Predicting the binary black hole population of the Milky Way with cosmological simulations}

\author[A. Lamberts et al.]{A. Lamberts$^{1}$\thanks{E-mail: lamberts@caltech.edu}, S. Garrison-Kimmel$^{1}$, P. F. Hopkins$^1$, E. Quataert$^2$,  J. S. Bullock$^3$,
\newauthor  C.-A. Faucher-Gigu{\`e}re$^4$, A. Wetzel$^5$,  D. Kere{\v s}$^{6}$, 
K. Drango$^{1}$, and R. E. Sanderson$^{1}$
\\
$^{1}$ Theoretical Astrophysics, California Institute of Technology, Pasadena, CA 91125, USA\\ 
$^2$ Department of Astronomy and Theoretical Astrophysics Center, University of California, Berkeley, CA 94720, USA\\
$^3$ Center for Cosmology, Department of Physics and Astronomy, University of California, Irvine, CA 92697, USA\\
$^4$ Department of Physics and Astronomy and CIERA, Northwestern University, Evanston, IL 60208, USA\\
$^5$ Department of Physics, University of California, Davis, CA 95616, USA\\
$^6$ Department of Physics, Center for Astrophysics and Space Science, University of California at San Diego, 9500 Gilman Drive,\\ La Jolla, CA 92093, USA
}

\date{Accepted XXX. Received YYY; in original form ZZZ}

\pubyear{2018}

\begin{document}
\label{firstpage}
\pagerange{\pageref{firstpage}--\pageref{lastpage}}
\maketitle

\begin{abstract}

Binary black holes are the primary endpoint of massive stars. Their properties
provide a unique opportunity to constrain binary evolution, which remains
poorly understood. We predict the main properties of binary black
holes and their merger products in/around the Milky Way. We present the first
combination of a high-resolution cosmological simulation of a Milky Way-mass
galaxy with a binary population synthesis model in this context. The
hydrodynamic simulation, taken from the FIRE project, provides a cosmologically
realistic star formation history for the galaxy, its stellar halo and
satellites. During post-processing, we apply a metallicity-dependent
evolutionary model to the star particles to produce individual binary black
holes. We find that $7\times10^5$ binary black holes have merged in the model Milky
Way, and $1.2\times10^6$ binaries are still present, with a mean mass of 28
$\msun$. Because the black hole progenitors are strongly biased towards low
metallicity stars, half reside in the stellar halo and satellites and a
third were formed outside the main galaxy. The numbers and mass distribution of
the merged systems is broadly compatible with the LIGO/Virgo detections. Our
simplified binary evolution models predicts that \textit{LISA} will detect more
than 20 binary black holes, but that electromagnetic observations will be
challenging. Our method will allow for constraints on the evolution of
massive binaries based on comparisons between observations of compact objects
and the predictions of varying binary evolution models. We provide online data
of our star formation model and binary black hole distribution.
\end{abstract}


\section{Introduction}

The global properties of compact objects (CO) in the Milky Way (MW) provide crucial information on the star formation history of the Galaxy as well as on stellar evolution. Broadly speaking, stars born with mass $M<8\msun$ evolve into white dwarfs (WD), those of $M\simeq8-20$ evolve into neutron stars (NS), and those above $\simeq 20\msun$ turn into black holes (BH; e.g. \citealt{Fryer99_BHmass}, but also see \citealp{Sukhbold16_progeitormasses}), though stars between $\simeq 120-250 \msun$ may undergo a pair instability supernova that leaves no remnant \citep{Fryer:2012}.  Aside from systems that have undergone mergers or left the Galaxy due to BH kicks \citep{Janka13_kicks}, the number of compact remnants quantifies past star formation.  The localization of COs within a galaxy may also be indicative of the progenitor's formation conditions (lookback time, local environment, and metallicity). The mass distributions, orbital properties and/or proper motion of the COs can inform us on stellar evolution and explosion mechanisms. In this paper, we provide detailed predictions for the expected binary black hole (BBH) population and their merger products in the MW, as well as some  observational properties.

The current ($z=0$) population of  BHs is particularly important as BHs evolve from the most massive stars, whose short lives mean that we can only directly observe the population that formed within the last $\sim20$~Myr. In particular, BHs provide unique information on the initial mass function of massive stars, which are key drivers of galactic evolution through chemical enrichment, stellar winds,  ionizing radiation and their final explosions \citep{Muratov15_galaxywinds,Geen15_feedback}. Observations indicate that most, if not all massive stars form in binary systems \citep{Sana:2012}, with recent work suggesting that binaries are even more ubiquitous for lower metallicity stars \citep{Badenes17_multiplicity}. As such, the properties of stellar black holes can also inform us about crucial phases of binary evolution such as supernova kicks and mass transfer (see \citealt{Postnov14_review} for a recent review on compact binary formation). 

Unfortunately, the inventory of stellar mass black holes in the MW is far from complete. So far, the only confirmed systems are found in X-ray binaries, where the BHs manifest themselves through accretion of material from a companion star (see \citealt{Casares17_XBreview} for a recent review). About 60 of these systems have been detected around low-mass companion stars, and a handful around massive stars \citep{Corral-Santana16_BHXRB}. The latter are  systems formed within the last 20 million years and possible progenitors to BBHs. The astrometric mission \textit{GAIA} could detect more than ten thousand BHs around stellar companions \citep{Breivik17_BHGAIA,Mashian17_BHGAIA}. Based on the observed BBH merger rate from \citet{LIGO:2016_rate}, \citet{Elbert17_BHcount} estimate that there could be up to 100 million BHs in the MW. So far, however, no BH has been observed in a binary with another compact object in the MW. 

Similarly, there have been no firm detections of stellar mass BHs without a stellar companion in the MW.  Such black holes are not expected to emit electromagnetic radiation unless they are accreting from a dense environment \citep{Agol02_XraysBH,Maccarone05_radioBH}. Future hard X-ray surveys or radio observations with the \textit{Square Kilometer Array (SKA)} could lead to the first such detections, if the accretion rates and radiative efficiencies are high enough \citep{Fender15_IBH,Corbel15_IBH}.  \citet{Ioka17_TeVBH} further suggest that BHs formed out of merged BBHs should have a high spin and may thus produce gamma-ray emission in a jet. Year-long microlensing events with no visible lens have been tentatively attributed to BHs in the galactic bulge \citep{Wyrzykowski10_OGLE_MACHOS,Wyrzykowski16_OGLE}.

Gravitational waves may be the most promising means of detecting BBHs in the MW. The first direct detection of GWs came from the merger of two stellar mass BHs (GW150914) via the Laser Interferometry Gravitational Wave Observatory (LIGO; \citealp{LIGO:2016_main}), with a handful of similar detections following. Several studies propose to distinguish binary evolution channels using observational properties of compact object mergers from gravitational waves
\citep{Belczynski02_DCO,Voss03_BPS,Belczynski08_DCO,Dominik12_DCO,Stevenson15_BPS,Mapelli17_Illustris}.  With LIGO/Virgo, we observe the very last moments (of order of a few seconds or less) before the BBH merges, the merger itself, and the ringdown of the newly formed (and more massive) BH.  Given the very short signal in comparison to the very long inspiral time of a BBH, the likelihood of detecting one of these events in the MW is close to zero.

The high masses of GW150914, $M_1=36_{-4}^{+5} \msun, M_2=29\pm 4\msun$ suggests a low-metallicity progenitor binary \citep{Dominik:2013,Belczynski08_DCO}, with $Z\lesssim 0.1\Zsun$ \citep{Belczynski:2016} \A{although $Z\lesssim 0.5\Zsun$ may be possible for very massive progenitors \citep{LIGO:2016_implications,Eldridge_BPASS_BBH}}. Based on an analytic model, we showed in \citet{Lamberts16_GW150914} that the progenitors of GW150915 most likely formed either relatively recently in a dwarf galaxy (stellar mass $\lesssim10^7\msun$), or around the peak of cosmic star formation ($\simeq$ 10 Gyrs ago) in a galaxy that would now resemble the MW (stellar mass $\sim10^{10}$--$10^{11}\msun$). \citet{Mapelli17_Illustris} and \citet{Schneider17_GameshBH} found similar results when combining a binary evolution model with the Illustris simulation and with a high-resolution dark matter-only simulation, respectively.  While lower mass BBHs can be formed out of higher metallicity progenitors, BBHs are strongly biased towards sub-solar metallicity environments, as the amount of mass loss from stellar winds scales with metallicity \citep{Belczynski10_massBH,Dominik:2013}. As such, we expect the BBH population of the MW  to have a different spatial distribution than the overall stellar mass.

Current predictions for the binary compact object population of the MW are based on simplified models for its star formation history, metallicity and morphology. The MW is often approximated by a spherically symmetric bulge and a disk with a characteristic scale height (e.g. \citet{Nelemans01_WD,Ruiter2010_WD,Liu14_LISA}). The star formation rate in the disk is typically assumed to be constant over time and to occur at fixed metallicity (solar).  The stellar halo is rarely included (except in \citealp{Belczynski2010_LISA}, where the halo is assumed to form in a single burst 13~Gyr ago with a single metallicity).  The inaccuracies resulting from these approximations are compounded by the uncertainties in the binary evolution models, such that it is extremely difficult to constrain binary evolution through comparisons between predictions from these simplified Galaxy models and observations.

These simplifications motivate the present analysis, where we instead apply a population synthesis model to the star formation history of a cosmological, hydrodynamic simulation of a MW-mass galaxy, first presented in \citet{Wetzel2016}. Although the simulation does not specifically aim to reproduce the exact morphology of the MW, it includes a cosmologically realistic star formation and merger histories, satellite population, and stellar halo, and a  self-consistent metal enrichment and gas exchange with the circumgalactic and extragalactic media.  An exact prediction for the BBH distribution in the MW requires observational constraints on the ages and metallicities of the entire stellar population of the MW, and would be particularly sensitive to the oldest (and therefore faintest) stars. Unfortunately, this detailed information will remain out of reach for some time, especially outside the disk of the Galaxy. In the meantime, our technique should yield a binary black hole population that is statistically consistent with that of the MW.

The \textit{Laser Interferometer Space Antenna (LISA)}, a space-based mission led by ESA and scheduled to launch in the mid 2030's, will provide the first view of the double compact object population in the MW. With its 2.5 million~km arms, the interferometer will be mostly sensitive to gravitational wave frequencies $10^{-4} \lesssim f_{GW}\lesssim 10^{-1}$ Hz. One of \textit{LISA's} main science goals is a first inventory of very short orbit double compact objects (DCO) in the MW. WD binaries with orbits shorter than an hour are expected to vastly dominate the signal \citep{Nelemans01_WD,Ruiter2010_WD}, as they stem from more common low mass stars and their formation likely has limited metallicity dependence.  However, analytic estimates \citep{Seto16_BBHLISA,Christian17_BBH_LISA} and binary population synthesis models combined with a multi-component model for the Galaxy \citep{Belczynski2010_LISA,Liu14_LISA} predict that at most a few BBH may be detected. However, \citet{Belczynski2010_LISA} predict roughly $8.0\times 10^5$ BBH in the Galaxy in their model A, where binaries survive the common envelope occurring during the Hertzprung gap. They predict roughly $5.6\times 10^{5}$ binaries in model B, where common envelope mass transfer during the Hertzprung gap results in a stellar merger.  In both cases, they predict most of the BBHs will be in the disk, a quarter of the systems in the bulge, and a negligible contribution from the halo. We will show that this arises directly from their neglect of more complex chemical and star formation history in the Galaxy. By applying a binary population synthesis model (similar to model B of \citealt{Belczynski2010_LISA}) to a more realistic stellar population (both in terms of spatial distribution and age-metallicity space), we predict that a few tens of BBH will be detected in the MW with \textit{LISA}.

\begin{figure}
\centering
\includegraphics[width = .45\textwidth]{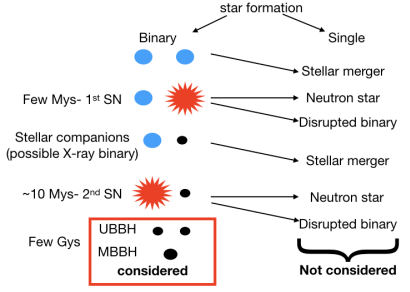}
\caption{Possible endpoints of binary massive star formation. In this paper we only focus on unmerged binary black holes (UBBH) and merged binary black holes (MBBH), in red in the bottom left. We discard black holes from single stellar evolution, stellar mergers or binaries unbound by supernova kicks. We also do not consider black holes with white dwarf, neutron star or stellar companions. The above picture is a cartoon view to clarify the systems considered here and is not intended to be a precise representation of binary evolution.}
\label{fig:evol} 
\end{figure}

In this paper, we provide a detailed view of the black hole population resulting from binary black holes systems in a MW-mass galaxy.  We only focus on systems that survived BBH formation and are either unmerged binary black holes (UBBH) or merged binary black holes (MBBH). We do not consider black holes from single star formation, black holes resulting from binary systems disrupted by supernova kicks, black holes with a stellar companion, or single black holes formed after a stellar merger. We specifically consider UBBH+MBBH black holes and do not present earlier phases of binary evolution where BHs may have stellar companions emitting electromagnetic radiation. Fig.~\ref{fig:evol} summarizes the systems considered in this work.

We combine a high resolution cosmological simulation of a MW-mass halo \citep{Wetzel2016} with a binary population synthesis model  (\S\ref{sec:methods}). We show how the progenitors of the BHs compare with the global population of stars and determine the properties of the BH population (\S\ref{sec:results}). We present observational properties of both merged (i.e. currently single BHs) and unmerged systems (binary BHs), both with gravitational waves and electromagnetic methods (\S\ref{sec:detection}).  We discuss the importance of a detailed model of the MW (\S \ref{sec:discussion}) and conclude (\S\ref{sec:conclusion}).

\section{Methods}\label{sec:methods}
This paper emphasizes the importance of a realistic model for the star formation history for MW-mass galaxies with respect to previous studies of binary compact objects in the MW. We first present the key numerical and physical parameters of the simulation and show the physical quantities that are the most relevant to this study (\S\ref{sec:MW_model}). We then present the binary evolution model we use (\S\ref{BPS_model}) as well as our computation of GW emission (\S \ref{GW_model}). We then explain how all these aspects are combined together (\S\ref{sec:full_model}).

\subsection{A realistic model of a Milky Way-mass galaxy}\label{sec:MW_model}

The inputs to our binary evolution model (the ages, metallicities, and positions of star particles) are primarily drawn from the \textbf{m12i} FIRE-2 simulation, also known as ``Latte'' \citep{Wetzel2016}. The simulation has an initial gas particle mass of 7070 M$_{\odot}$. The Latte simulation is part of the Feedback in Realistic Environment (FIRE; \citealp{Hopkins:2014_FIRE}) project\footnote{\url{http://fire.northwestern.edu}}, specifically run using the improved ``FIRE-2'' version of the code from \citet[][for details, see Section~2 therein]{Hopkins2017fire2}. The simulations use the code GIZMO \citep{Hopkins2015gizmo},\footnote{\url{http://www.tapir.caltech.edu/~phopkins/Site/GIZMO.html}} with hydrodynamics solved using the mesh-free Lagrangian Godunov ``MFM'' method. For the gas, both the hydrodynamic and gravitational (force softening) resolutions are fully adaptive down to 1 pc. The simulations include cooling and heating from a meta-galactic background and local stellar sources from $T\sim10-10^{10}\,$K.  Star formation occurs in locally self-gravitating, dense, self-shielding molecular, Jeans-unstable gas. Stellar feedback from OB and AGB star mass-loss, type Ia and II supernovae, and multi-wavelength photo-heating and radiation pressure is  directly based on stellar evolution models. Chemical enrichment stems from type Ia supernova \citep{Iwamoto99_SNyields}, core-collapse supernova \citep{Nomoto06_SNyields}, and O and AGB star winds \citep{VandenHoek99_AGByields,Marigo01_AGByield,Izzard04_AGByield}. All the binary evolution models are included during post-processing, and the hydrodynamic simulation does not explicitly include binary effects.

The FIRE simulations reproduce the observed mean mass-metallicity relation both for stars and star forming gas, between $z=0$ and $z=3$ \citep{Ma:2016} down to a stellar mass of $10^6\msun$.  The simulations here include the subgrid-scale numerical turbulent metal diffusion terms described in \citet{Hopkins2017fire2}, which have almost no dynamical effect at the galaxy mass scales considered here \citep{Su2017_metaldiff}, but produce better agreement with the internal metallicity distribution functions observed in MW satellite galaxies \citep{Escala2017_dmdf}.

Our main analysis is based on galaxy {\bf m12i} (from \citealp{Wetzel2016}, though we analyze a re-simulation with turbulent metal diffusion first presented in \citealp{Bonaca2017}), chosen to have a relatively ``normal'' merger history, but we also consider a lower-resolution version of {\bf m12i} as well as two different galaxies {\bf m12b} and {\bf m12c} \citep{Hopkins2017fire2} at the same mass scale. {\bf m12i} shows metallicity gradients \citep{Ma17_metalMW} and abundances of $\alpha$-elements (Wetzel et al, in prep.) in the disk that are broadly consistent with observations of the MW. Its global star formation history is consistent with the MW (see \citet{Ma17_metalMW} for illustrations) although its present day star formation rate of 6$\msun$ yr$^{-1}$ is somewhat higher than observed in the Milky Way. The satellite distribution around the main galaxy in {\bf m12i} presents a similar mass and velocity distribution as observed around the Milky Way and M31, down to a stellar mass of $10^5\msun$, though the simulation does not contain an equivalent of the Large Magellanic Cloud; the most massive satellite is comparable to the Small Magellanic Cloud. Outputs from the simulation and corresponding mock Gaia catalogs are available online \footnote{https://fire.northwestern.edu/data/ and  http://ananke.hub.yt } \citep{Sanderson2018}.

The simulation produces a catalog of \A{roughly 14 million particles of about} $\simeq 7000 \msun$ in mass.\footnote{Whenever we refer to the simulation, we use the words star, particle and star particle interchangeably.} For each particle, the quantities of interest here are its formation time $t_*$, metallicity $Z$, and position at $z=0$ and at $t_*$. To determine the accretion rates of each BH associated with a star particle,  we also recover the properties of the closest surrounding gas particle and assign it to the star particle. We consider only particles within 300 kpc of the center of the main galaxy. This is slightly larger than the Virial radius of the galaxy and allows us to largely sample the halo, satellites and streams while remaining unaffected by the boundaries of the high resolution region.

The simulations assume a $\Lambda$CDM cosmology with $\Omega_{\Lambda}$ = 0.728, $\Omega_m$ = 0.272, $\Omega_b$ = 0.0455, h = 0.702, $\sigma_8$ = 0.807, and n$_s$ = 0.961 \citep{Planck16_cosmo}. All metallicities are defined with respect to the solar metallicity, set to $\Zsun=0.02$.

\subsection{Binary evolution model}\label{BPS_model}
We consider binary evolution in the field, neglecting possible GW sources from N-body stellar dynamics in globular clusters or other formation channels \citep{Rodriguez:2015,OLeary16_dynamicalBBH,Mapelli16_runaway_BBH}, including Pop III stars \citep{Kinugawa:2014}. We focus on the current standard picture of massive binary evolution and neglect alternate channels based on chemically homogeneous evolution \citep{Mandel:2016,Marchant:2016}.  Throughout the paper, quantities refer to BH properties and we use superscript `$*$' to refer to properties of the progenitor stars in ambiguous cases (such as the mass).

\begin{figure*}
\centering
\includegraphics[width = .95\textwidth]{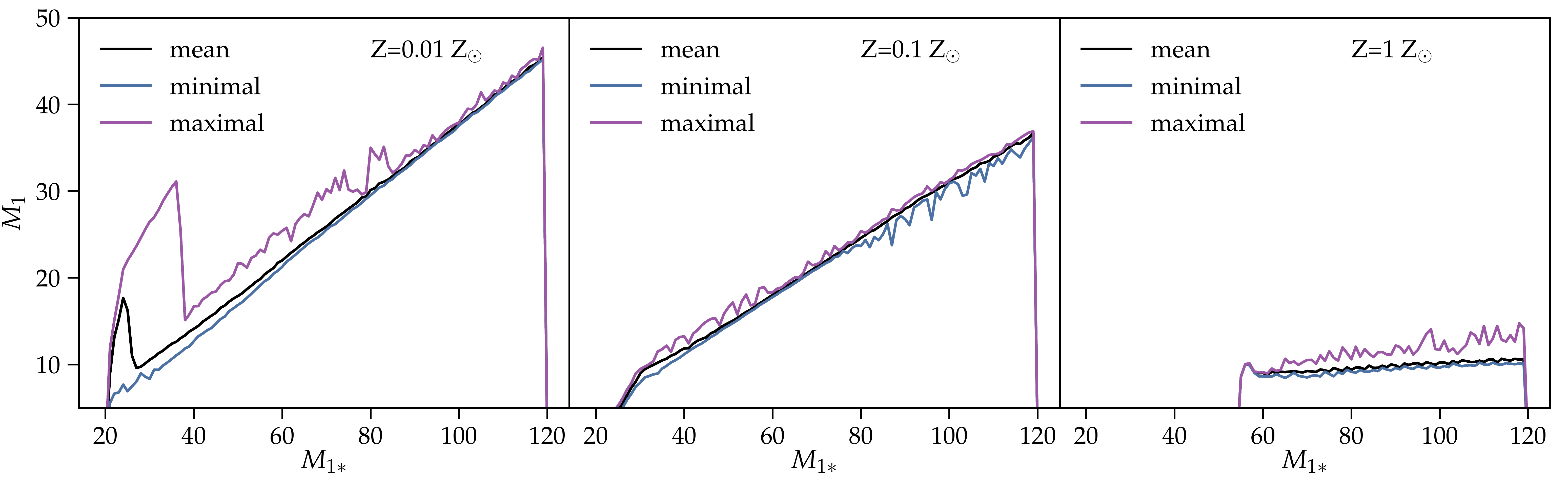}
\caption{Mean (black) and extremal masses of the primary black hole as a function of the primary progenitor mass for increasing metallicity for the binary evolution model presented here.} 	
\label{fig:Mini_MBH} 
\end{figure*}

As in \citet{Lamberts16_GW150914}, we use the binary stellar evolution code (BSE, \citealp{Hurley:2002}), with modifications for massive binaries. Rapid binary population synthesis codes like BSE are based on prescriptions for the stellar evolution and interactions based on the initial  masses and metaliicty of the stars. Such codes allow for a computationally efficient exploration of a wide range of parameters, but can miss important effects in the stellar structure and mass transfer \citep{Eldridge_BPASS_BBH}. 

Currently, the main uncertainties in the evolution of massive binary stars are their  mass-loss rates (especially for low-metallicity stars), the outcome of the common envelope interactions, the effect of SN kicks and the remnant masses. We use the metallicity-dependent prescription for the mass loss rates from \citet{Belczynski10_massBH}. We use the  simplified prescription from \citet{Belczynski08_DCO} for the remnant mass, which neglect details about the stellar structure \citep{Eldridge_04_SNprogenitors,Sukhbold16_progeitormasses}. We use the model from \citet{Dominik:2013} for the BH initial kicks, which are drawn from a Maxwellian distribution that peaks at 265~km~s$^{-1}$. The kicks are then reduced according to the amount of material that falls back after core collapse (i.e. by a factor $M_{BH}/M_{NS}\simeq M_{BH}/1.4\msun$, such that more massive BHs experience smaller kicks) consistent with the analysis by \citet{Mandel_16_BHkicks}. For example, a $14\msun$ BH (roughly the median mass of the BHs in our binaries) is given a typical kick of $\sim25$ km s$^{-1}$. This is lower than values quoted in~\citet{Repetto_12_BHkicks}, based on the locations of low-mass X-ray binaries in the Milky Way. The latter assume the binaries were formed in the midplane of the galaxy and find that kick velocties comparable with neutron star kicks are necessary in order to reach their current location. Based on our simulations, we find that most of the binaries are in the halo or the thick galactic disk. As such, their locations can be explained  independently of natal kicks and solely based on the cosmological assembly of the Galaxy and its large scale dynamics.
For the common envelope mass transfer, we use the so-called $\alpha$-formalism \citep{1984ApJ...277..355W} using the common envelope efficiency $\alpha=1$ and the envelope binding energy is determined according to the evolutionary stages of the stars. When common envelope occurs during the Hertzprung gap (between core hydrogen burning and shell hydrogen burning), we assume that a stellar merger occurs, as the boundary between the core and envelope is too smooth to stop the inspiral (\citealp{Ivanova:2004}, referred to as model B in \citealp{Belczynski:2007}). Globally, we assume that half of the mass lost by the donor during Roche-lobe overflow is accreted by the secondary. Fig.~\ref{fig:Mini_MBH} shows the resulting mean, minimal and maximal mass of the primary black hole as a function of the primary progenitor mass for $=0.01, 0.1, 1\Zsun$. \A{In general binary interactions lead to a wider range of possible black holes masses, than might be expected from a single star of the same initial mass.}

We create 13 different samples from the BSE model with metallicities logarithmically spaced between $Z=0.005\Zsun$ and $1.6\Zsun$ ($\Zsun\equiv 0.02$)\footnote{We use $Z=5\times 10^{3}, 10^{-2}, 1.6\times 10^{-2}, 2.5\times10^{-2},4.0\times 10^{-2},6.3\times 10^{-2}, 0.10, 0.16, 0.25, 0.40, 0.63, 1.0, 1.6$ times the solar metallicity.}, which are the limits currently allowed in BSE. 
For each metallicity, we build a statistical sample of binaries, with primary masses $M_{1*}$ between 20 and 120 $\msun$ following a Kroupa IMF \citep{Kroupa2001}.  
The secondary masses $M_{2*}$ are drawn assuming that the mass ratios are uniformly distributed, and the initial period distribution is taken from \citet{Sana:2012} (power law in log space with exponent $p=-0.55$).  As (single) stars above 120 $\msun$ are likely subject to pair-instability supernova, which does not leave any remnant~\citep{Heger:2003}, we choose not to expand our upper limit.  We begin with a  thermal distribution for the eccentricities ($f(e)\propto e$), which favors high-eccentricity systems. While the initial conditions of the binary evolution (masses and orbital parameters) are somewhat uncertain, \citet{Demin15_initialcond} showed that these uncertainties only affect the final results by at most a factor 2. We set the binary fraction to unity; all numbers can be directly rescaled to lower values. The number of binaries simulated with BSE depends on the metallicity, and is adjusted in order to ensure that our star particles are matched to a smooth distribution of BBHs.  In practice, we require at least 3500 BBHs per metallicity bin, which requires an initial sample of approximately 30 000 binaries at $Z=0.005\Zsun$ but roughly $5\times10^7$ at the highest metallicity.

\begin{figure*}
\centering
\includegraphics[width = .95\textwidth]{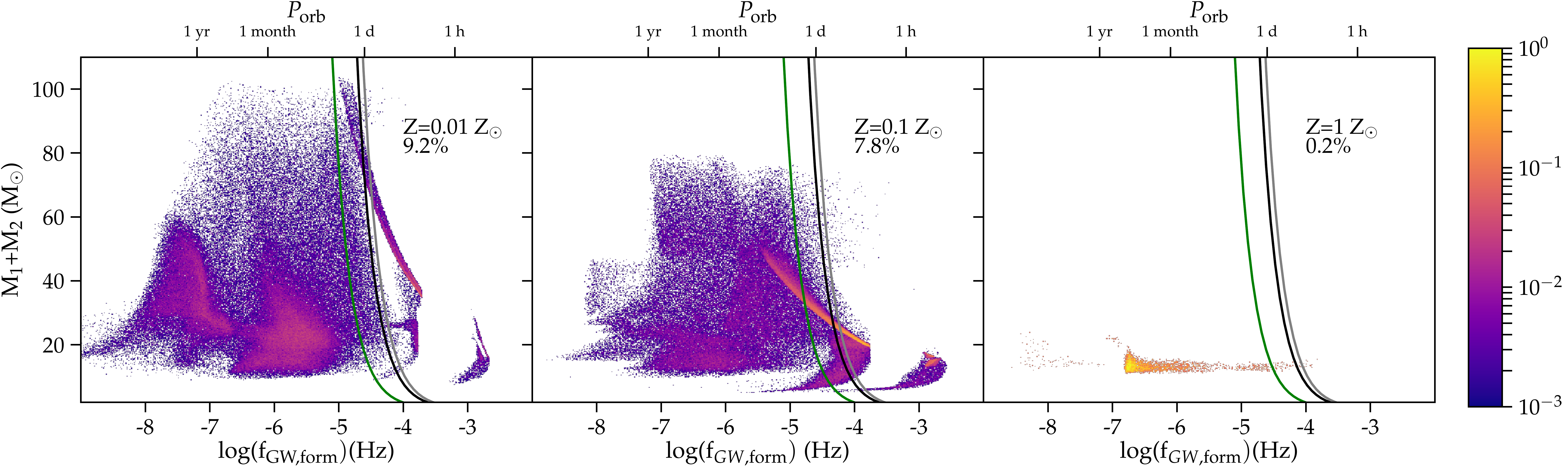}
\caption{Properties of BBHs at their formation times for different stellar metallicities. We show the total binary masses ($M_1+M_2$) versus initial orbital period ($P_{\mathrm{orb}}$) or equivalently  the gravitational wave frequency of an equivalent circular orbit $f_{\mathrm{GW}}$ (the orbits are not necessarily circular, but this is intended only to guide the reader). The colors are normalized to the total number of BBH formed in each metallicity bin while the percentage value shows the percentage of massive binaries that evolve into BBHs.  The black/grey lines across the plot shows the maximal initial period for circular systems to be able to merge, assuming they formed at the beginning of the universe or at $z=1$, respectively. The green line shows the minimal period for a merger within a Hubble time for BBHs with initial eccentricity $e>0.7$, which is the case for about 1 per cent of the systems at all metallicities. Most systems left of the black line cannot have merged by the present day. }	
\label{fig:fini_Mc} 
\end{figure*}

At most 8 per cent of the initial binaries result in BBHs (for the lowest metallicity). About two thirds of the binaries have companions that are too low mass (initial $M_{2*}\lesssim 20 \msun$) to allow the formation of a second black hole. Among the massive enough binaries, about a third will undergo stellar mergers (see Fig.~\ref{fig:evol}). This is consistent with the estimates of \citet{deMink14_stellar_merger} that about 10$\%$ of massive stars on the main sequence are merger products. Roughly another third of the binaries that would be massive enough to yield a BBH will instead be disrupted during the one of the supernova explosions; the remaining third will result in a BBH. At metallicity beyond $\simeq 0.3\Zsun$, BBH creation is available to less than one per cent of stellar binaries, due to mass loss from stellar winds.

Fig.~\ref{fig:fini_Mc} shows the total mass mass $M_{\mathrm{tot}}=(M_1+M_2)$ and orbital period $P_{\mathrm{orb}}$ at the formation of the BBH according to our binary evolution model. Assuming circular orbits, we indicate the corresponding gravitational wave frequency (twice the value of the orbital frequency). These plots shows that BBH formation is rare at solar metallicity, and limited to $M_{\mathrm{tot}}\leq 30$, in agreement with \citep{Belczynski:2016,Eldridge_BPASS_BBH}. The lowest frequency systems stem from binaries that never interacted. At $Z\le 0.3 \Zsun$, BBH formation occurs for about $7\%$ of  more massive binaries, but only the lowest metallicity model routinely produces binaries with $M_{\mathrm{tot}}>60\msun$, as stars at higher metallicities lose most of their mass in strong winds due to higher opacities in their atmosphere.

More specifically, for the lower metallicity models, several channels to BBH formation appear. The systems with the shortest initial orbit (or highest frequency) result from initial stars with $M_{*} \leq 50\msun$, a stellar mass ratio $q$ close to unity and a wide enough initial orbit to avoid stellar mergers during the common envelope phase. The narrow strip of binaries with orbits of order of a day typically have $M_2\ge M_1$ and result from more massive stars with initial mass fractions close to unity. These systems have undergone important mass transfer from the primary to the secondary.  The other binaries come from initially closer orbits, with stars with $0.3\leqslant q \leqslant 0.8$ and $M_{1*} \leqslant 35 \msun$. Systems with periods of order a month or larger typically come from equal mass binaries at higher masses. 

The main output of the BPS is a list of 3500 ``sample'' binaries for each metallicity. For each binary, we record its initial stellar masses and orbital properties, the formation time of the BBH $t_{\mathrm{form}}$ with respect to the formation of the progenitors and its masses and orbital properties. In the next section, we explain the gravitational wave properties of a given binary.

\subsection{Gravitational wave emission}\label{GW_model}
After the BBH is formed, the binary evolves only via gravitational wave radiation, gradually shortening the orbit. To assess the BBH population of the galaxy at any given point in time $t$, we first need to determine whether a binary with given properties has already merged. The time to coalescence is given by e.g. \citet{Maggiore}
\begin{equation}
T_m=9.829\,\mathrm{Myr}\, \left(\frac{T_0}{1\,\mathrm{hr}\,}\right)^{8/3} \left(\frac{\msun}{M_1+M_2}\right)^{2/3}\left(\frac{\msun}{\mu}\right)F(e_0) ,
\end{equation}
where $\mu$ is the reduced mass of the BBH, $e_0$ the initial eccentricity of the binary and $F$ a function depending on the orbital evolution of the system ,which is equal to unity for circular binaries. (see Eq. 4.137 in \citet{Maggiore}.).

If the binary has not yet merged, its semi-major axis $a$ and eccentricity $e$, which determine its GW emission, evolve according to \citep{Peters63}
\begin{eqnarray}\label{eq:orb_evol}
\frac{de}{dt}&=&-\frac{304}{15}\frac{G^3\mu (M_1+M_2)^2}{c^5a^4 }\frac{1}{(1-e^2)^{5/2}}\left(1+\frac{121}{304}e^2 \right)\\ \nonumber
\frac{da}{dt}&=&-\frac{64}{5}\frac{G^3\mu (M_1+M_2)^2}{c^5 a^3}\frac{1}{(1-e^2)^{7/2}}\left(1+\frac{73}{24}e^2+\frac{37}{96}e^4\right).\\ \nonumber
\end{eqnarray}
For eccentric sources, GWs are emitted over a range of harmonics of the orbital frequency, while only the second harmonic emits for circular orbits.  The total energy loss can be significantly higher than for circular orbits, resulting in faster inspirals. Even though many binaries are eccentric at birth, we find that the orbits are close to circular ($e<0.15$) by the present day. As such, we include eccentricity for the orbital evolution, but compute the present day GW emission assuming circular orbits.
For circular orbits, the frequency and characteristic strain of the GW at a given distance $d$ of the source are given by
\begin{eqnarray}
f_{GW}&=&\frac{1}{\pi}\sqrt[]{\frac{G(M_1+M_2)}{a^3}}\\ \label{eq:freq_GW}
h_c&=&\left(\frac{32}{5}\right)^{1/2}\frac{(M_cG)^{5/3}}{dc^4}\pi^{2/3} f_{GW}^{7/6}, \label{eq:strain}
\end{eqnarray}
where the chirp mass is given by $M_c=(M_1M_2)^{3/5}(M_1+M_2)^{-1/5}$.

\subsection{From a MW model to GW emission}\label{sec:full_model}
The core of this paper is the combination of a binary population synthesis model (\S\ref{BPS_model}) with a cosmological model for a MW-mass galaxy (\S\ref{sec:MW_model}) and its star formation history as a function of localization and metallicity to determine the GW emission (\S\ref{GW_model}).   Here we describe how these three aspects are combined during post-processing of the simulation.

We first bin the star particles from the \textbf{m12i} simulation into the same 13 metallicity bins as our BPS model. Stars with metallicity below/above the range covered in the BPS model are assigned to the first/last bin. Each star is then randomly assigned a binary from the BPS model at the corresponding metallicity.  This effectively associates black hole masses $M_1$, $M_2$, an orbital period $P_0$ and eccentricity $e_0$ of the BBH at formation with each star. We also keep track of the formation time of the BBH with respect to the formation of the progenitor stars $t_{\mathrm{form}}$.  

To obtain the present-day distribution of BBH, we determine the time $dt=t_H-(t_{\mathrm{form}}+t_*)$ over which to evolve the binary in order to reach $t_H$, the present-day age of the Universe.  We evolve all the BBHs forward during $dt$ according to Eq.~\ref{eq:orb_evol} and their initial orbits and masses. For each BBH, we then have the present-day orbital parameters if it has not yet merged (from GW evolution); and spatial localization (from the simulation) and can determine its gravitational wave properties and possibilities for electromagnetic detection.

In this method, we have assumed that each star particle can be uniquely associated with a single BBH. In practice, the star particles have a mass $\simeq 7000 \msun$ and represent an IMF-averaged group of stars. From the Kroupa IMF, we find that there are about 12 binaries with $M_{1*}\geq 20$ in such a star particle. For each metallicity, we can determine the expected number of BBHs according to the fraction of massive stars that effectively turn into a BBH (see percentages in Fig.~\ref{fig:fini_Mc}). Effectively, we find that star particles with $0.025\leqslant Z \leqslant 0.4$ typically form one BBH, stars with lower metallicity create between 1 and 2, and stars with higher metallicity rarely make BBHs. As such, we weigh all of our mock statistical BBHs with the expectation value of the number of BBH associated with an IMF-averaged stellar population of the same mass, age, and metallicity as the simulation star particle. We perform a polynomial fit of the fraction of BBHs as a function of metallicity and use this to extrapolate the probability of producing a BBH for systems beyond our initial metallicity range.\footnote{Due to the strong dependence on metallicity (as shown in Fig.~\ref{fig:fini_Mc}), this extrapolation produces a negligible number of additional BBHs and therefore has a minimal impact on our results.} This weighting is accounted for in all the results presented here. 

Although we include BH natal kicks to determine the initial orbits of the BBH in our BPS model, we do not assign these same kicks to our particles and the location of the BH is set by the location of the star particle they stem from. As we discuss in \S\ref{sec:discussion}, we estimate that this approximation does not impact the main results of our study.

\section{Black holes in a MW-mass galaxy}\label{sec:results}
\begin{figure}
\centering
\includegraphics[width = .45\textwidth]{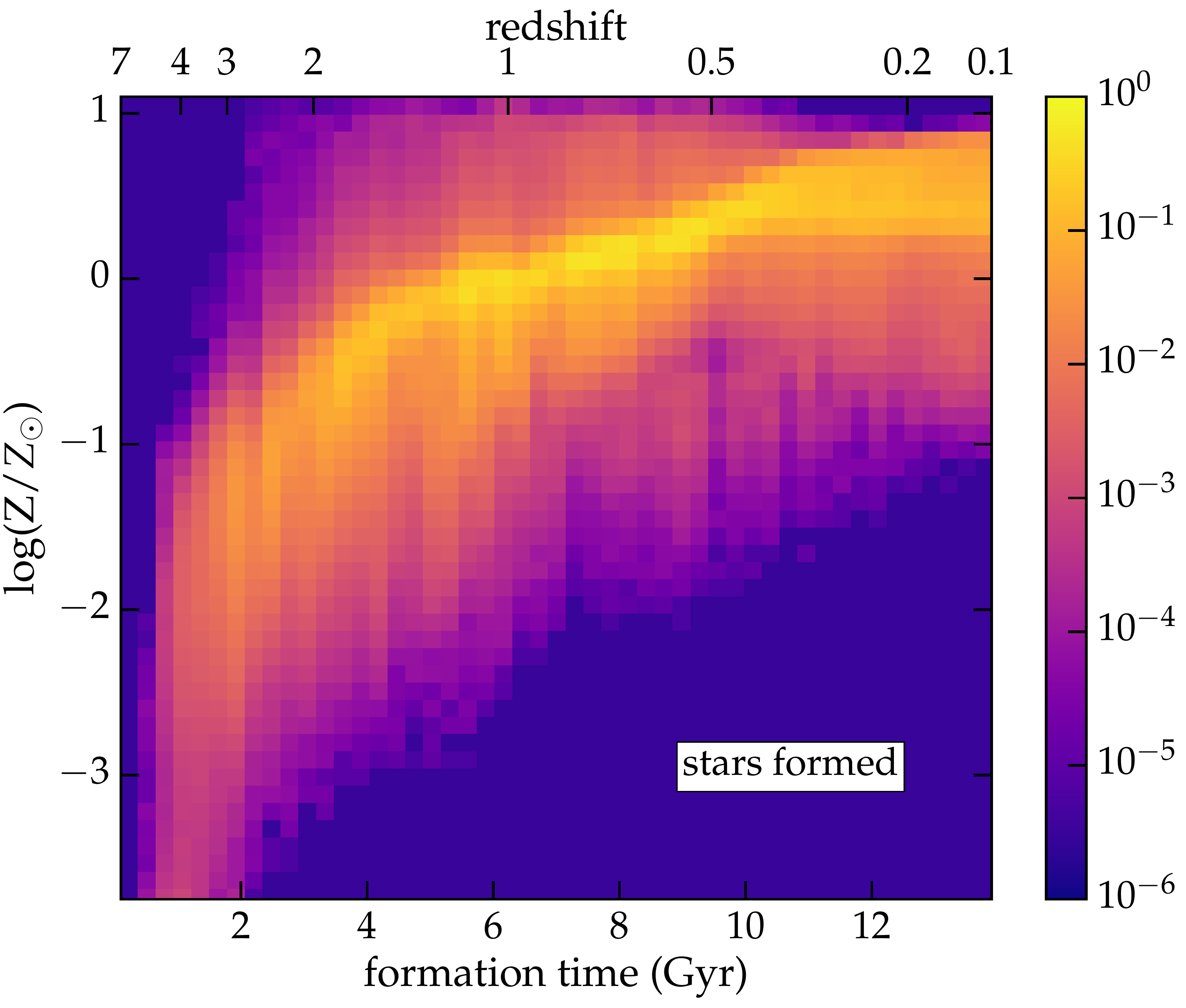}\\
\includegraphics[width = .45\textwidth]{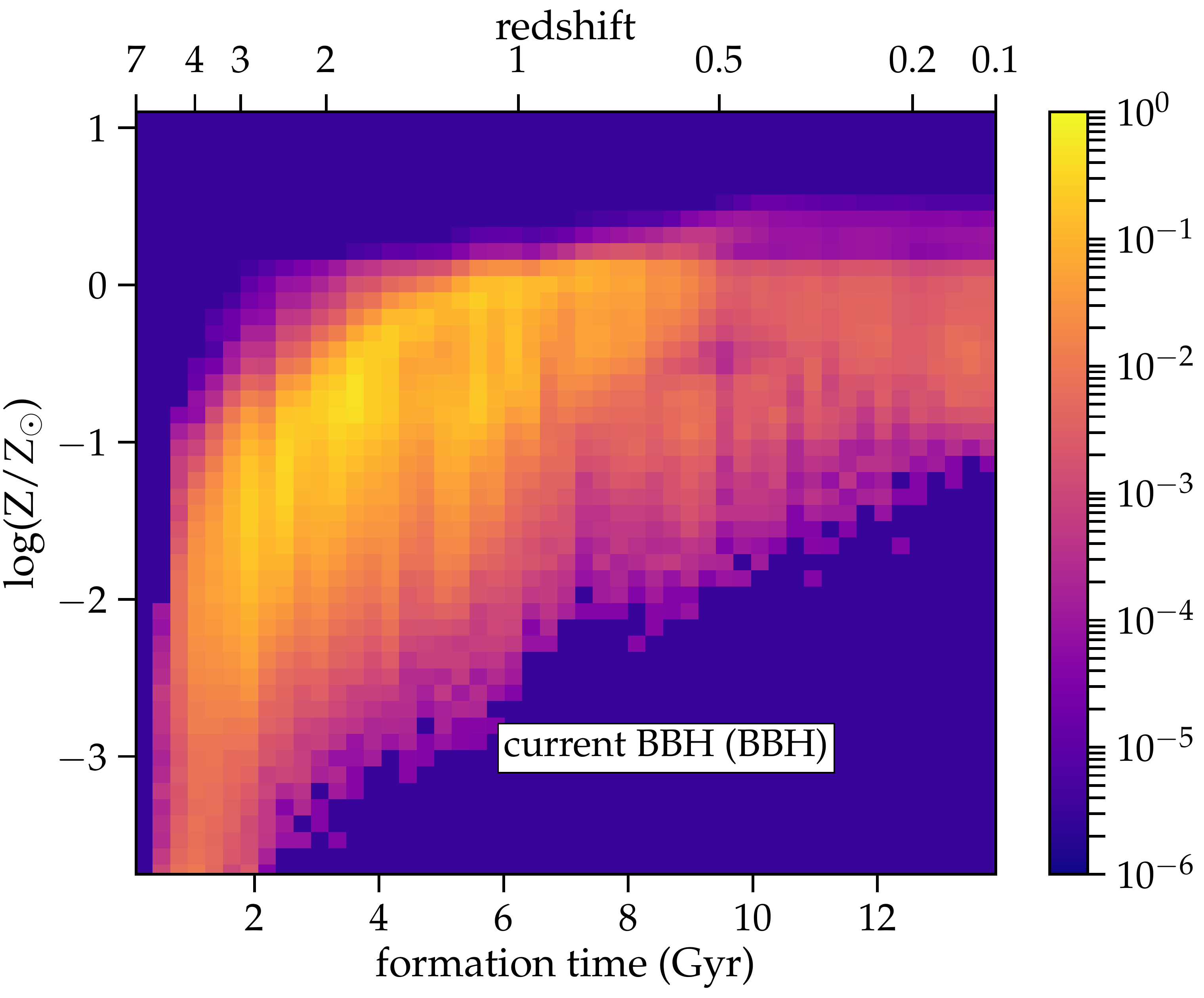}\\
\includegraphics[width = .45\textwidth]{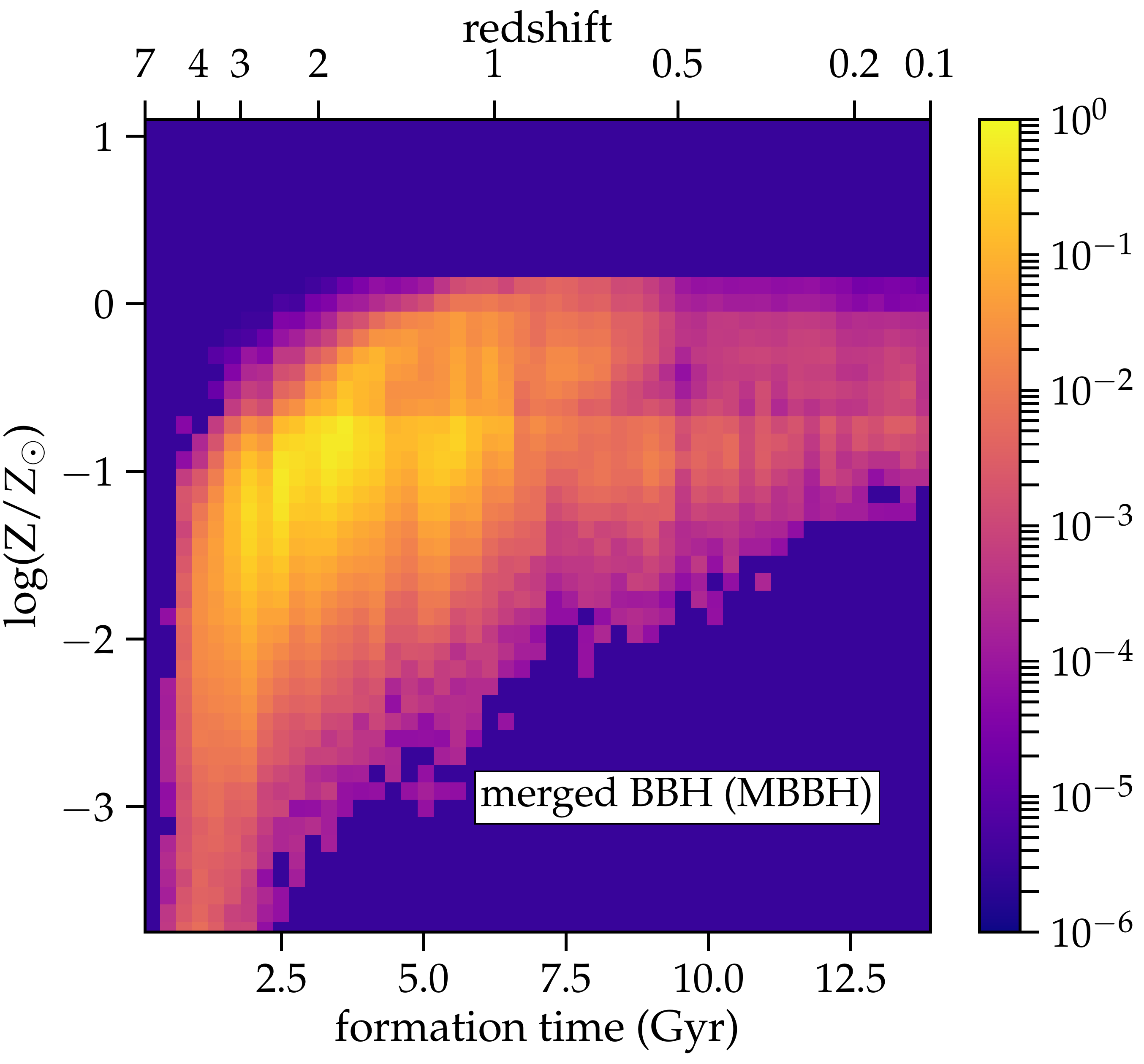}
\caption{Distribution of the formation time and metallicity of all the stars within 300 kpc of the center of the galaxy in the simulation (top), those stars that are progenitors of the currently unmerged-BBHs (middle) and the progenitors of the BBHs that have already merged (bottom). All distributions are normalized to unity.}	
\label{fig:SFR_Z_m12} 
\end{figure}

Fig.~\ref{fig:SFR_Z_m12} shows the distribution of formation time and metallicity of all the stars within 300 kpc (top) and the subset of stars that are UMBHs at  $z=0$ (middle) as well as systems that have already merged (MBBH, bottom).  The star formation extends from very early times to the present day, and is consistent with the global peak of star formation between 10 and 4 Gyr ago \citep{Madau14_cosmicSFR}. While the stellar metallicity globally increases over time, the scatter is important at all ages. The BPS model (\S\ref{BPS_model}) highlighted that the formation of BBHs very strongly depends on metallicity. While the bulk of recent star formation is significantly super-solar, about a percent of the recent star formation is low enough metallicity to potentially allow BBH formation. Recent star formation in {\bf m12i} occurs at slightly higher metallicities than observations suggest is the case in the MW \citep{Mackereth2017:APOGEE}, but we emphasize that the drop-off in BBH formation with increasing metallicity is so steep that our predictions would only be marginally impacted by reducing the metallicity of late-time star formation by a factor of two.  However, this difference does somewhat depress the number of BBHs in the disk of our model MW. The data for this plot is provided at https://fire.northwestern.edu/data/. We also provide the list of binary black holes and their masses, present-day gravitational wave frequency and position in the galaxy.

The middle plot shows the progenitor stars of the UMBH population. The distribution is very limited above $Z\simeq 3\Zsun$, because BBH formation becomes extremely rare (see Fig.~\ref{fig:fini_Mc}). This means that in MW-mass galaxies, the bulk of the massive stars formed recently are unable to form BBHs. Star formation over the past 500~Myrs accounts for 3$\%$ of the total stellar mass but only $0.3\%$ of the mass in black hole binaries. Most of the currently present BBHs come from stars formed 8 to 10 Gyrs ago (between $z=1$ and $z=2$), when low metallicity star formation was the strongest. 

The distribution of progenitor stars of  MBBH is shown in the bottom panel.  Globally, the systems that have already merged by $z=0$ originate as lower metallicity stars, and thus trace older star formation. This confirms the results presented in \citet{Lamberts16_GW150914}. The upper limit on the metallicity is around $Z\simeq 0.2 \Zsun$ in the BPS model presented here, and the cutoff is even more drastic than for the unmerged systems.  This is because the few BBHs that are formed at higher metallicity typically have wide orbits and low chirp masses due to differences that arise during the course of stellar evolution, implying that they will never merge within a Hubble time (see Fig.~\ref{fig:fini_Mc}, and also delay time distributions in \citealp{Belczynski08_DCO,Dominik12_DCO,Lamberts16_GW150914}).  The exact cutoff is strongly dependent on the details of the BPS model, such as the mass loss through stellar winds, BH natal kicks and the outcome of common envelope evolution, and will likely be revised as our understanding of massive stellar binaries improves. However, an upper limit on the metallicity is likely to persist. Models accounting for strong mixing within the stars \citep{Mandel:2016,Marchant:2016} are also limited to low metallicity progenitors and are likely to stem from the same progenitor population.

\begin{figure*}
\centering
\includegraphics[width = .45\textwidth]{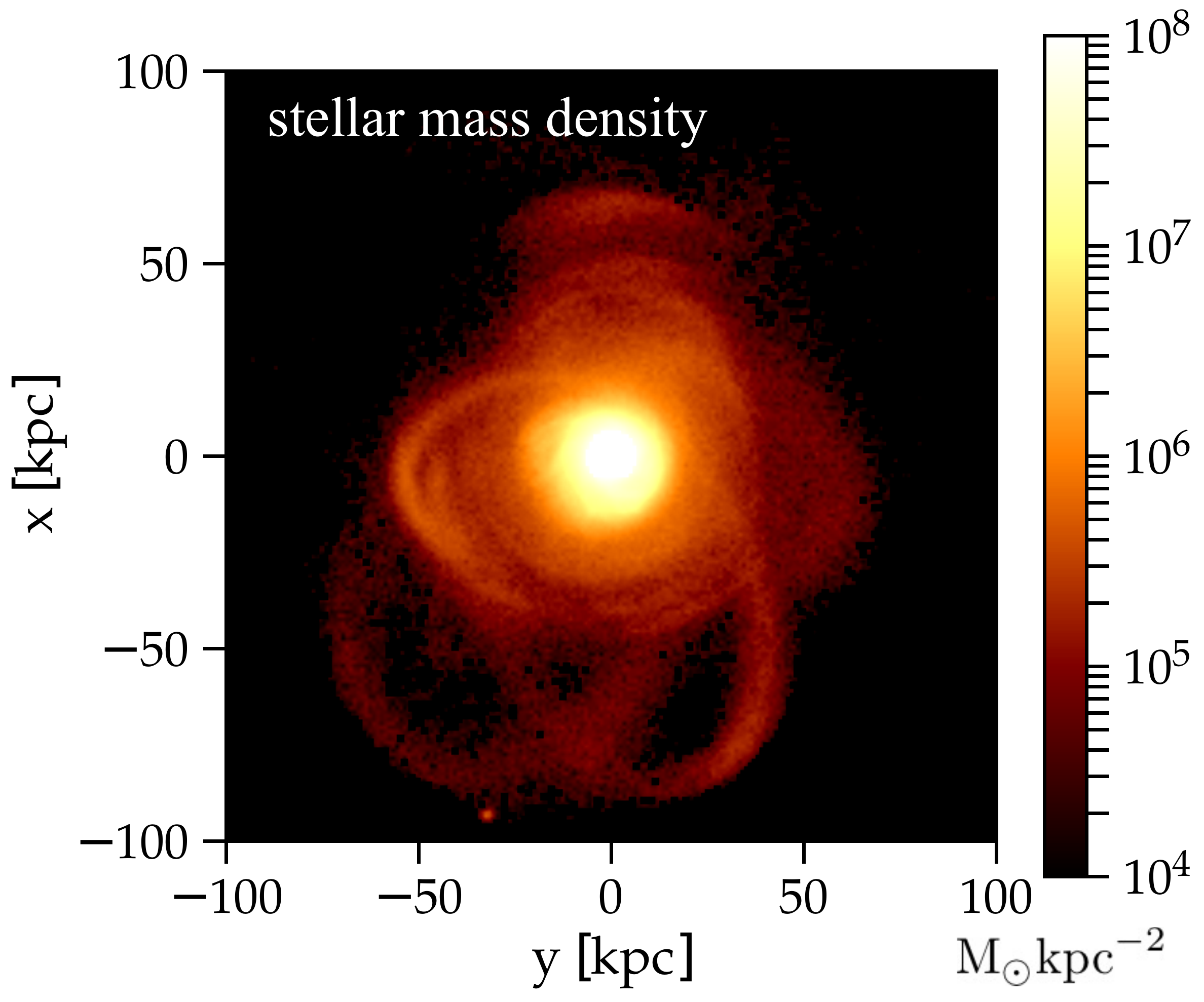}
\includegraphics[width = .45\textwidth]{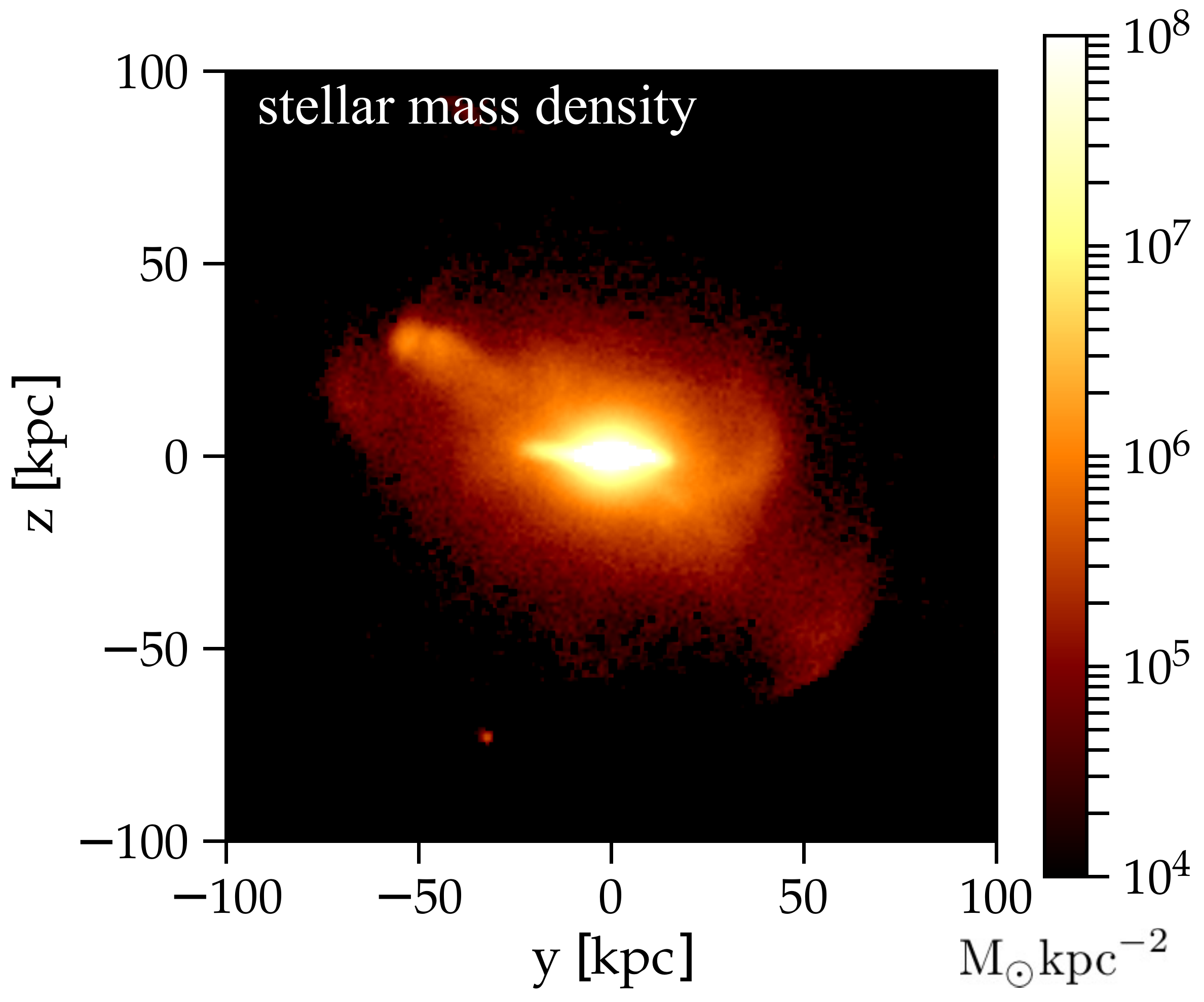}\\
\includegraphics[width = .45\textwidth]{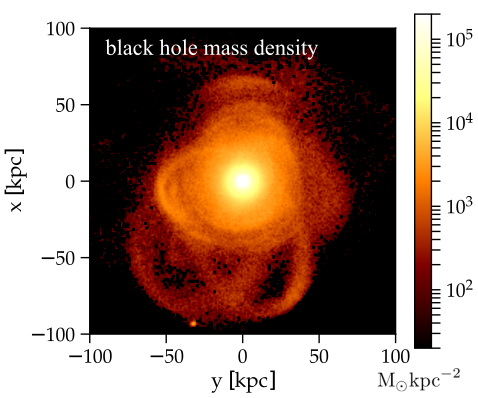}
\includegraphics[width = .45\textwidth]{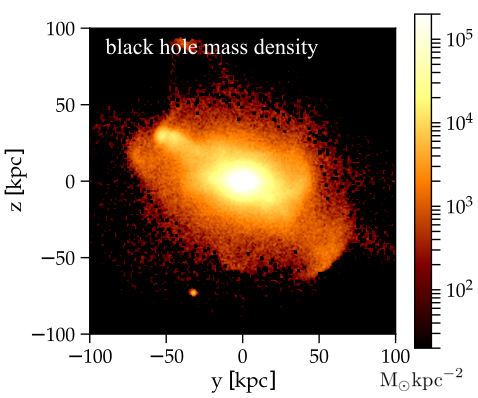}
\caption{Maps of the projected stellar mass density (top) and the isolated black hole mass density (merged and unmerged) (bottom) for a face-on (left) and edge-on view (right). The two rows show different absolute values (in units of Solar mass per kpc$^2$), but the dynamic range of the colors is identical, allowing for comparison between the two spatial distributions. The upper row represents the total stellar mass and is not a mock observation, which would prominently show young stars and include dust attenuation.  At smaller scales, galactic components such as the bulge and thin and thick disc are more prominent. The halo, streams and satellite galaxies are  overrepresented in the IBH maps owing to their lower metallicities, though we note that supernovae kicks, which we do not include in this work, may smear BBHs relative to the stellar streams and potentially eject them from the satellites. As described in \citet{Wetzel2016}, the cosmological simulation produces a realistic distribution of the MW satellites down to a stellar mass of $10^5\msun$.}
\label{fig:BBH_map} 
\end{figure*}

These plots show that UBBH/MBBH in MW-mass galaxies stem from a different population than most of the stars. As such, we can expect them to also have a different spatial distribution. Fig.~\ref{fig:BBH_map} shows maps of the distribution of stellar mass (top) and mass in (merged and unmerged) BBH for a MW-mass galaxy viewed face-on (left) and edge-on (right). The stellar mass distribution shows a zoomed-out view of a spiral galaxy with a central bulge and bright disk. The halo, which is more scarcely populated extends to about 40 kpc. Beyond, we find satellites and streams due to infalling satellites. In comparison, the BBH distribution is much less concentrated in the bulge and disk; instead, the halo is much more populated. Additionally, the satellites and streams over-produce BBHs with respect to their stellar content. Satellites are low mass galaxies (see \citealt{Wetzel2016} for a full description) with low metallicity star formation. Accordingly, these are prime sites for BBH production and mergers \citep{Lamberts16_GW150914,Elbert17_BHcount,Mapelli17_Illustris,Schneider17_GameshBH}. 

\begin{figure*}
\centering
\includegraphics[width = .45\textwidth]{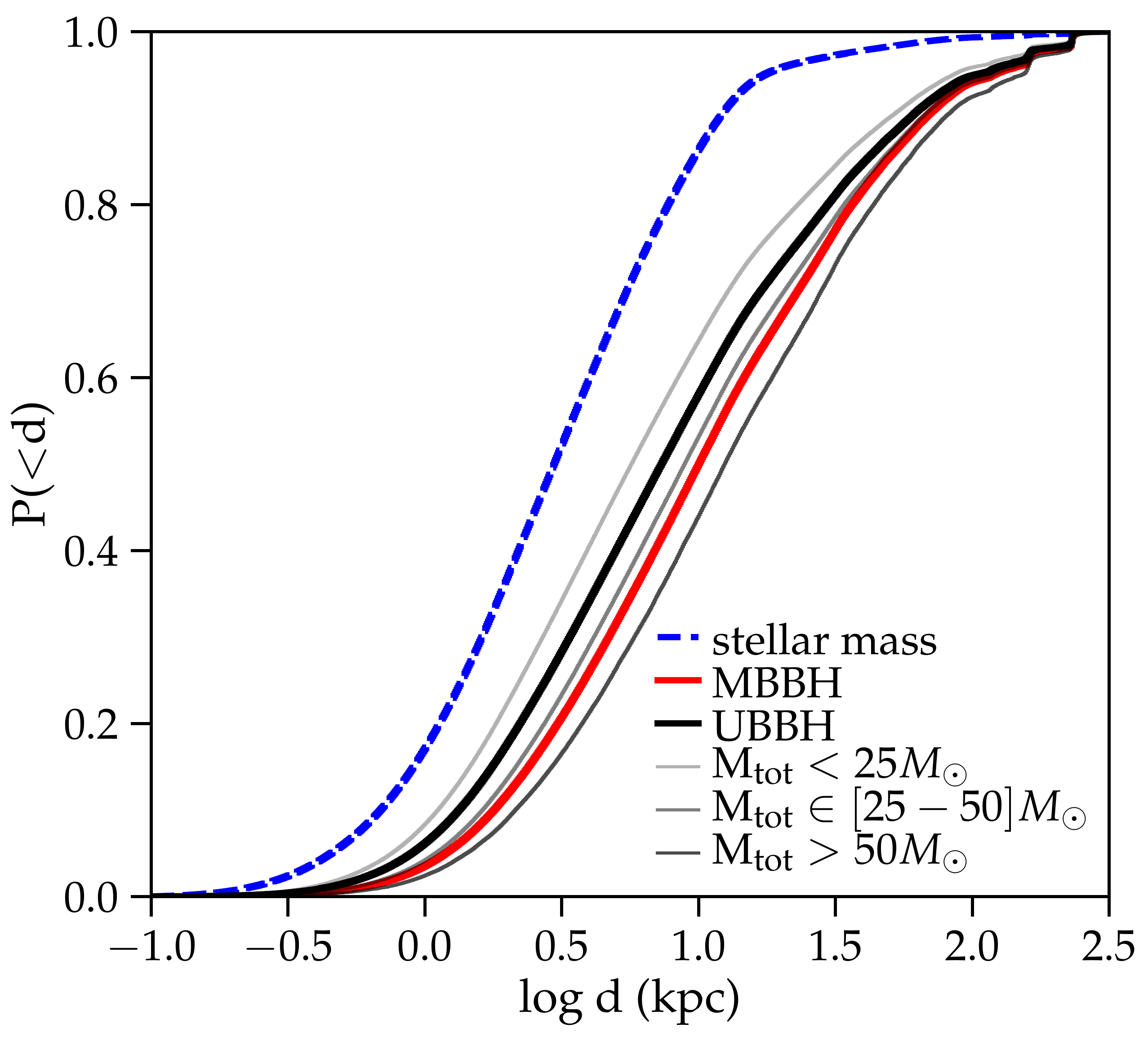}
\includegraphics[width = .45\textwidth]{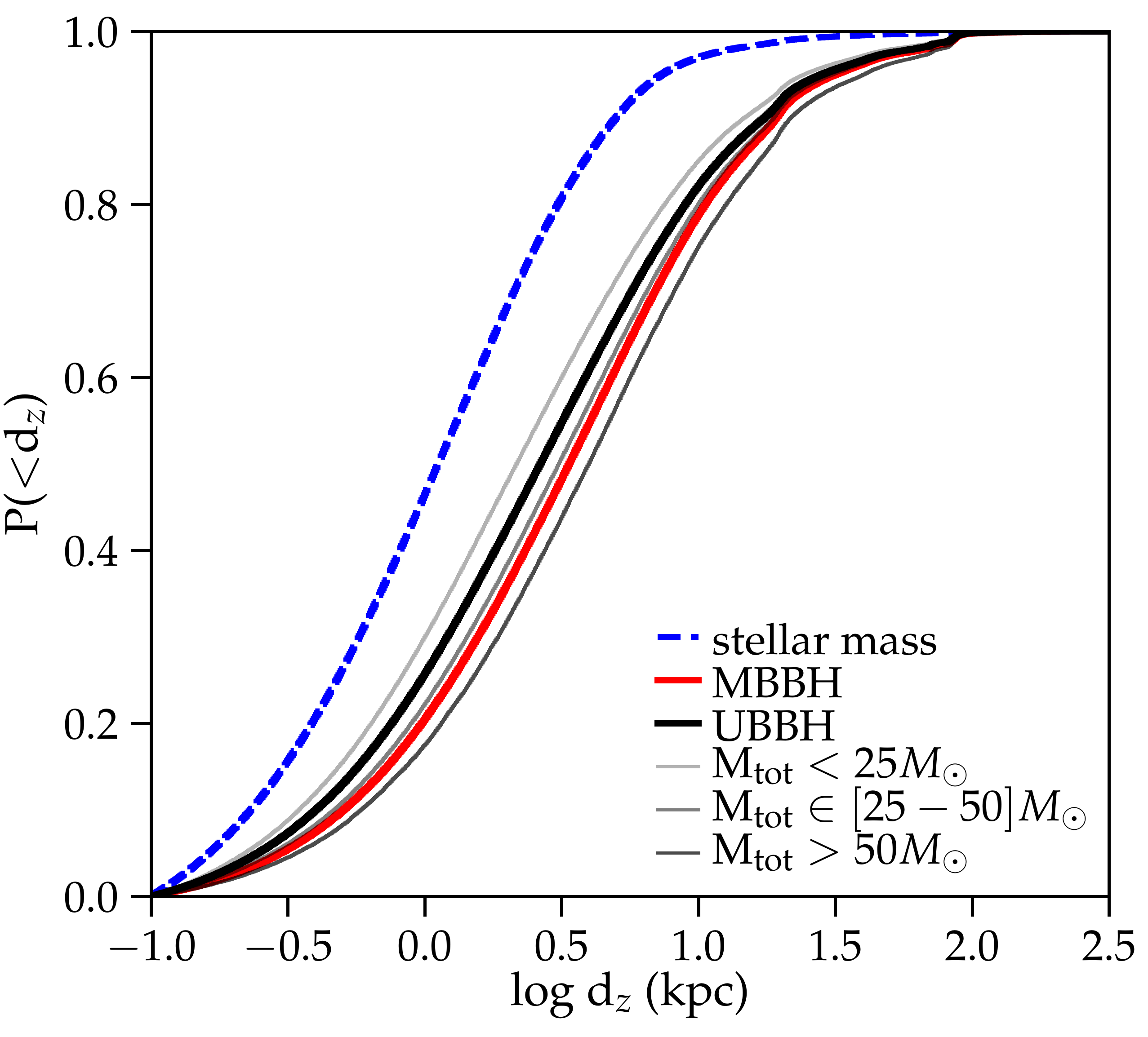}
\caption{Cumulative distribution function of the number of merged black holes (MBBH, red) and unmerged black holes (UBBH, black) as a function of distance to the galactic center (left) and vertical distance away from the plane of the galaxy (right). The distribution of the stellar mass is shown as a blue dashed line for comparison. For the merged systems, we distinguish different mass ranges with different grey scales. More massive binaries require lower metalicity and are further biased towards the halo and streams and satellites.}	
\label{fig:CDF_dist} 
\end{figure*}

Fig.~\ref{fig:CDF_dist} provides a quantitative description of the spatial distribution of the BHs in a MW-mass galaxy. Globally, the BBHs are preferentially on the outskirts of the galaxy. The effect is even stronger for the merged systems. About 60 per cent of the systems are located more than 10 kpc from the center of the galaxy, while 95 per cent of the stellar mass is concentrated within that distance. The right panel plots BBH counts as a function of distance above or below the plane of the disk of the galaxy.  Roughly 50 per cent of the BBHs are located at least 3~kpc above or below the disk mid-plane. The sudden increases in the cumulative distribution function at radii beyond 100 kpc indicate BBHs in satellite galaxies, which effectively contribute between 5 and 10 per cent of the systems. When considering the mass distribution of systems, we find that the respective dominance of the stellar halo and satellites is even stronger for the systems with total masses above $50 \msun$, 10 per cent of which lie beyond 50 kpc in satellites and streams, even though the latter contribute less than 1 per cent of the total stellar mass.

\begin{figure*}
\centering
\includegraphics[width =.4\textwidth]{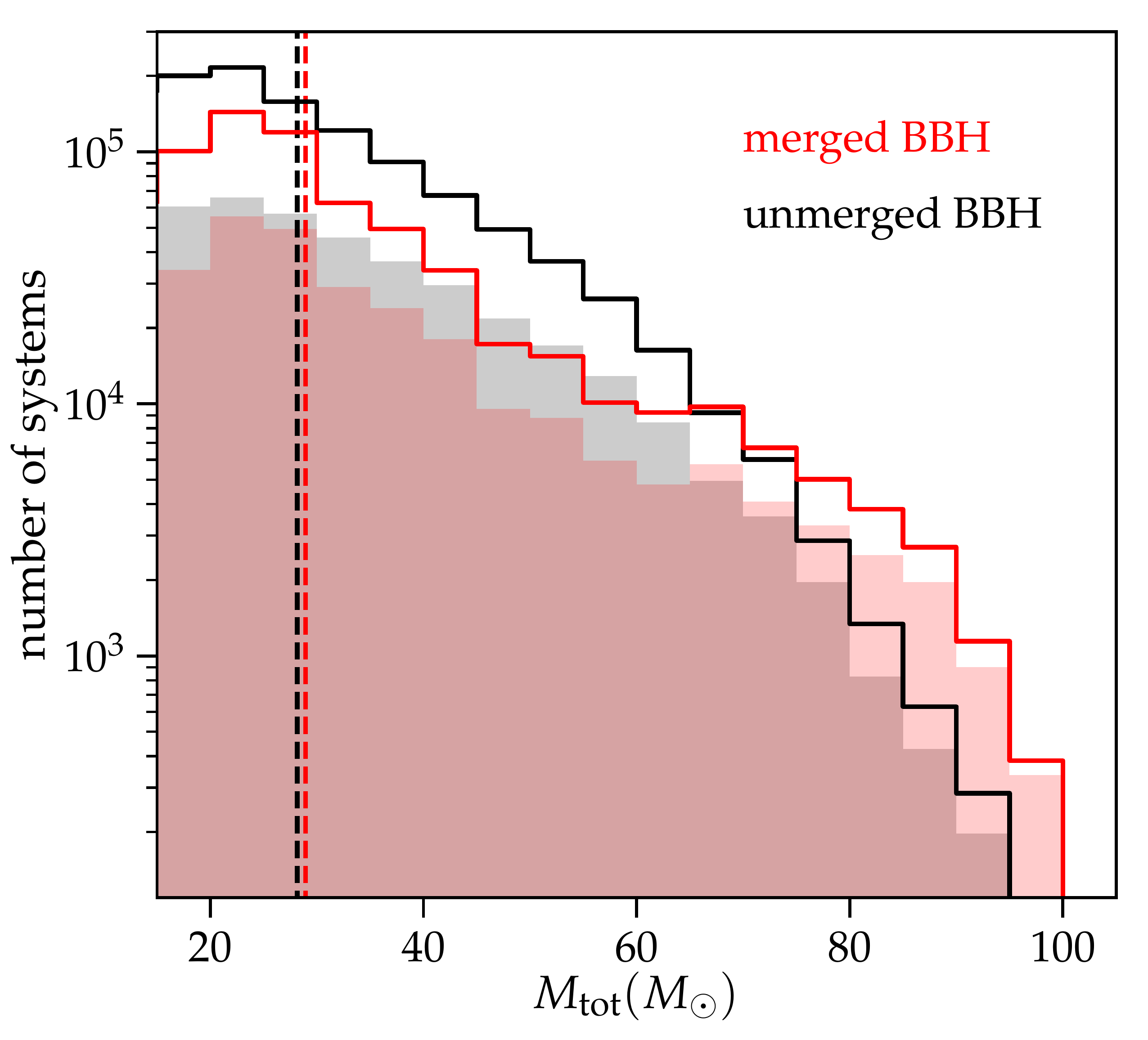}
\includegraphics[width = .4\textwidth]{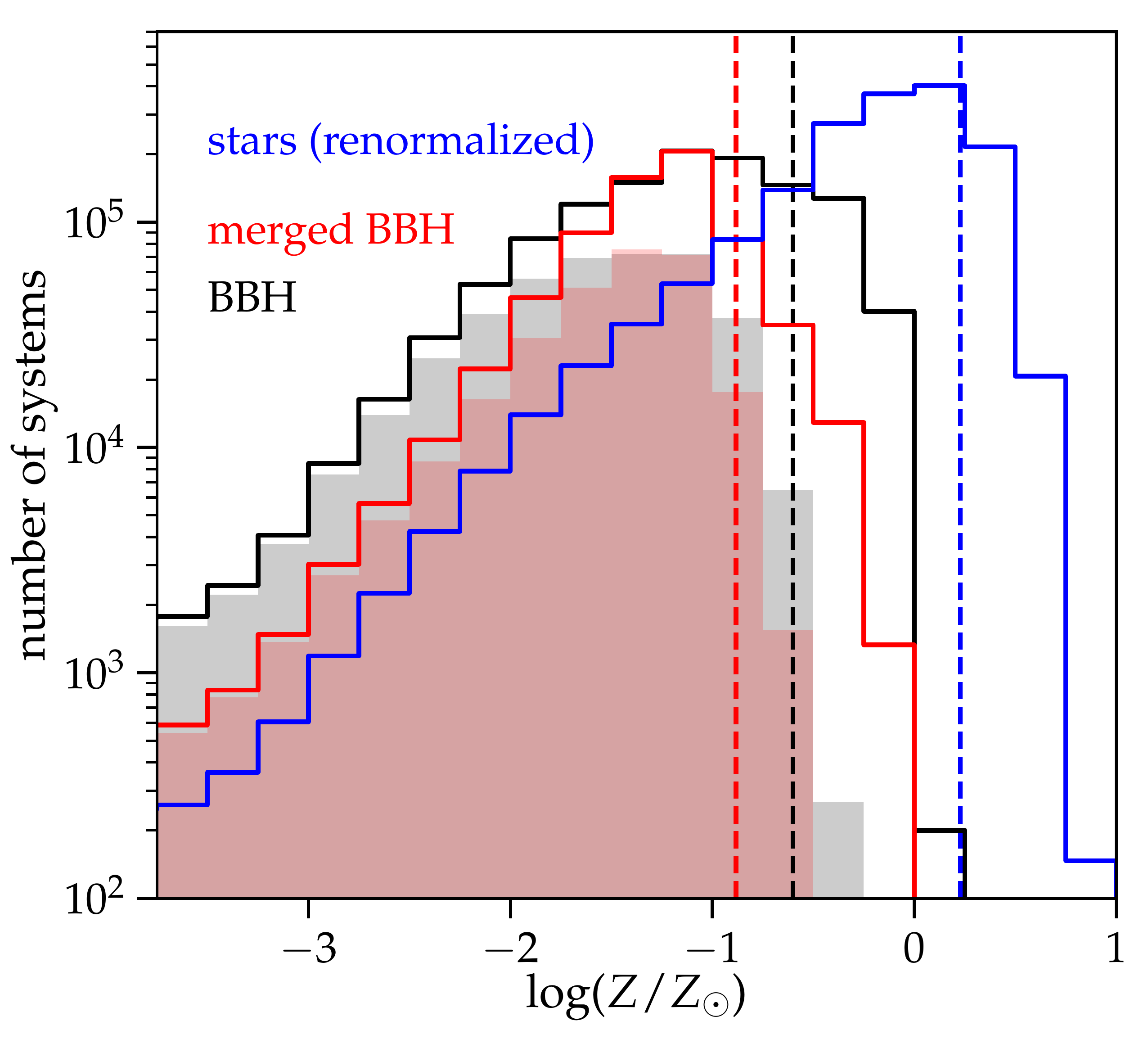}
\caption{Number of merged BBH (MBBH, red) and unmerged BBH (MBBH, black) as a function of total mass (left) and progenitor metallicity (right). The vertical dashed lines show the mean masses and metallicity of the merged and unmerged systems. The lines show the total number of systems while the shaded areas only represent systems formed outside of the main galaxy. The blue line on the right shows the stellar mass (divided by 5$\times 10^{4}\msun$) for comparison. Above $Z>0.01\Zsun$, the vast majority of the stars come from in-situ formation (not shown here).  The comparison between stars and BBH progenitors clearly shows that there are more total stars at high metallicity, but more efficient BBH formation at lower metallicity, producing the "peak" at $Z \simeq 0.1 \Zsun$.}
\label{fig:hist_BH_props}  
\end{figure*}

Fig.~\ref{fig:hist_BH_props} shows the distribution of merged and unmerged systems as a function of total mass of the BBH (left) and metallicity of the progenitor star (right). The total number of systems is shown with the solid lines. Out of a stellar mass of 7.3$\times 10^{10}\msun$ within 300 kpc, we find $6.8\times 10^5$ merged BBH systems  and $1.2\times 10^6$ unmerged systems. Out of the total current stellar mass, there is a total of $5.3 \times 10^7\msun$ in black holes from binary systems (merged or not merged). Assuming that the current stellar mass traces the total star formation we find that $0.07$ per cent of the stellar mass in a MW-mass galaxy turns into a black hole binary.  We remind the reader that we do not account for BHs with other companions or that have been kicked out of the initial binary. \A{\footnote{We note that some of the mass will be radiated away through gravitational waves as the binaries merge.} } We provide an estimate of the total number of black holes in a MW-mass galaxy in \S\ref{sec:discussion}.

The mean (median) mass is $28.9$ $(25.4)~\msun$ for the MBBHs. For the UBBHs the mean (median) total mass is $28.1$ $(24.7)~\msun$ for UBBHs. This is because most of the BBHs come from progenitors with $0.03 \geq Z/\Zsun \geq 0.2 $, where systems with wider obits, which have not merged yet, have more massive black holes (see Fig.~\ref{fig:fini_Mc}). Both for the merged and unmerged systems, massive binaries with total masses above $50\msun$ make up about 10$\%$ of the systems and systems above $80\msun$  make up less than $1\%$ of the systems.

The mean (median) metallicity of the progenitor stars is $0.13$ $(0.1)~\Zsun$ for the MBBHs and $0.25$ $(0.14)~\Zsun$ for the UBBHs. While extremely low metallicity progenitors (Z$< 0.01\Zsun$) are prime candidates for BBH formation (see \S\ref{BPS_model}), they only contribute about $5\%$ of the systems in MW-mass galaxies.  Conversely, progenitors formed with $Z>0.3\Zsun$ contribute about 30$\%$ of the unmerged systems and $10\%$ of the merged systems. In our model, progenitors with supersolar metallicity contribute less than 1$\%$ of the formed BBHs. The mean and median values are summarised in Tab.~\ref{tab:values}.

\begin{table}
 \caption{Summary of the mean and median properties of the merged and unmerged systems.}
 \label{tab:values}
 \begin{tabular}{ccccc}
  \hline
   &\multicolumn{2}{c}{$Z/\Zsun$}& \multicolumn{2}{c}{$M/\msun$}  \\ 
 \hline
   & merged &unmerged & merged & unmerged\\
   \hline
   mean & 0.15 & 0.26 & 26.5 & 28.4\\
   median & 0.11 & 0.20 & 22.5 & 24.5\\
   \hline
 \end{tabular}
\end{table}

The shaded area in both histograms shows the number of systems formed outside of the main galaxy. Stars are considered to have formed ex situ if their initial distance to the galactic center was more than 30 kpc (where "galactic center" refers to the center of the main progenitor of the z = 0 galaxy, at the time the stars formed).  Such systems have either merged into the galaxy by now (and are present in the bulge/halo), are in the process of being merged (and are present in streams) or are still present in the dwarf galaxy where they formed (and are now in satellites). In our MW model, less than 5 per cent of the currently present stellar mass is formed ex situ \citep{AnglesAlcazar17_assembly,Sanderson2017}. In comparison, we find that more than a third of the BHs (merged and not merged) come from ex-situ formation. For total BBH masses above 60$\msun$, only 40 per cent of the systems were initially formed within the main galaxy. This is because, more than 90 per cent of the stellar mass with $Z <0.01\Zsun$ originates outside the main galaxy, as does about half of the stellar mass with $Z < 0.1 \Zsun$. 

According to our combination of a cosmological model for a MW-mass galaxy and binary population synthesis, we find about 2 million black hole systems stemming from binary interactions, a third of which have merged by now. In the following section we describe the prospects of detecting these single and binary black holes with gravitational waves or electromagnetic signatures.

\section{Observational perspectives}\label{sec:detection}

\begin{figure}
\centering
\includegraphics[width = .45\textwidth]{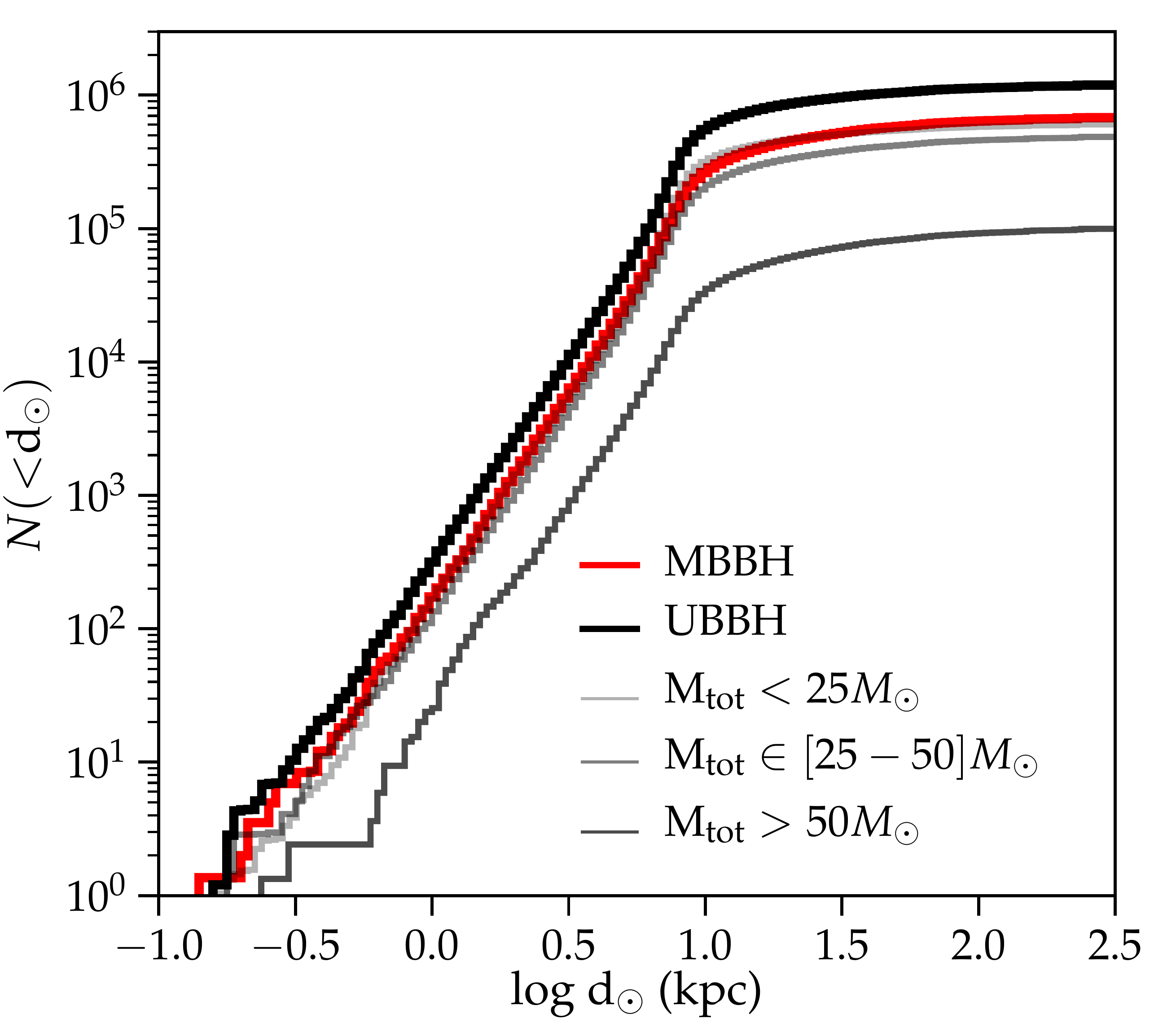}
\caption{Number of predicted merged (red) and unmerged (black) BBH  systems within a given distance of a randomly chosen point on the Solar Circle (i.e. 8~kpc from the galaxy center in the plane of the disk.)}
\label{fig:dist_to_sun} 
\end{figure}

Regardless of the detection method, the spatial distribution of the BBHs is a key in determining observational survey strategies.  Fig.~\ref{fig:dist_to_sun} shows the distance to the the sources (MBBHs and UBBHs) with respect to the Sun, i.e. at an arbitrary point along the Solar Circle:  on the disk mid-plane and 8~kpc from the center of the galaxy. About $300$ merged systems and $500$ binaries are present within a kpc. Unfortunately, the majority of both the merged and unmerged systems are located beyond 10 kpc.  Most of those are in the galactic halo,  off the plane of the disk (see Fig.~\ref{fig:CDF_dist}). Finding these sources would require the monitoring of a large fraction of the sky with a deep survey. 

Extragalactic mergers of BBH black holes have been detected with gravitational waves \citep{LIGO:2016_main} for $z<0.2$.  The local BBH merger rate inferred by LIGO is 12-213 Gpc$^{-3}$ yr$^{-1}$ \citep{LIGO_GW170608}. Assuming  $5\times 10^{-3}$ MW-mass galaxies per comoving  Mpc$^{-3}$ \citep{Baldry:2008}, this yields a galactic merger rate $6 \times 10^{-6} <R_{\mathrm{MW}} < 10^{-4}$ yr$^{-1}$, assuming all mergers occur in MW-mass galaxies. This makes a BBH merger very  unlikely in the MW \citep{LIGO:2016_rate}. At much lower frequencies, \textit{LISA} is currently the only planned GW detector that may detect BBHs before they merge. \textit{LISA} will be able to detect cosmological BBHs just a few years before they merge within the LIGO band \citep{Sesana16_binaries}, and it may also be able to detect BBHs in the MW or its nearby satellites thousands of years before they merge \citep{Belczynski2010_LISA}. Given the negligible likelihood of a merger occurring within the MW itself, we focus here on the latter possibility.

\begin{figure}
\centering
\includegraphics[width = .45\textwidth,trim= 0 0 15cm 0cm ,clip]{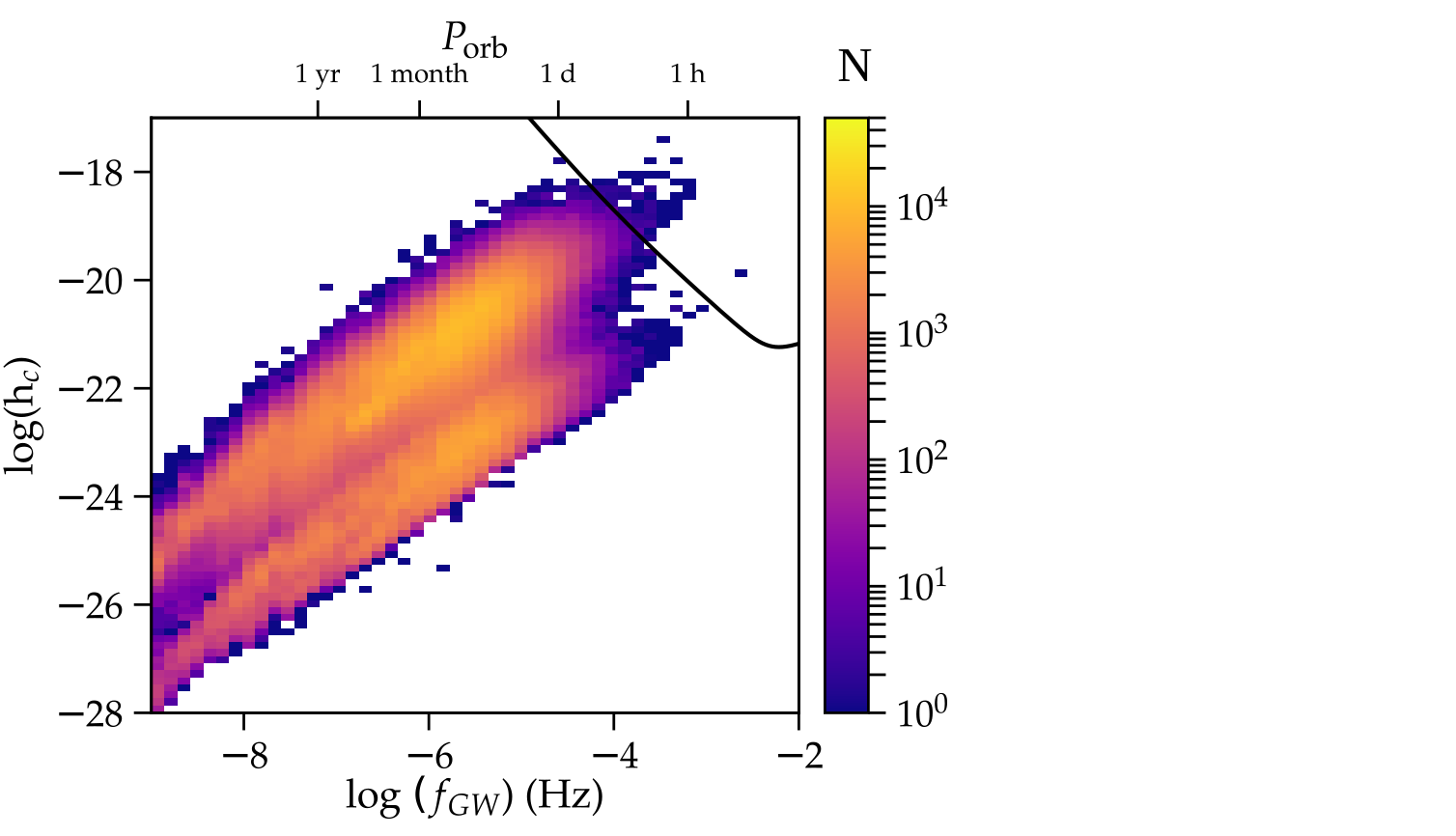}
\caption{Characteristic strain and GW frequency of the BBHs in our simulation assuming 4 years of observations. The black line in the upper right corner shows the current expected sensitivity curve of the \textit{LISA} mission -- only binaries above the line may be detectable. The color shows the number of systems in each bin.}	
\label{fig:f_vs_strain} 
\end{figure}
Fig.~\ref{fig:f_vs_strain} shows the gravitational wave frequency and characteristic strain expected from each binary after four years of observations in our model galaxy (Eqs.~\ref{eq:freq_GW}). It shows two distinct populations, with similar frequency distributions but different median values for the strain. The upper population is exclusively composed of binaries within the galaxy, while the lower group is composed of binaries within satellites. The ``loudest'' binaries ($h_c>10^{-20}$ ) have total masses above 40 $\msun$ and are all located within 30 kpc of the Sun, mostly within the disk. The black line in the upper left corner represents the expected noise curve for \textit{LISA} \citep{LISA,Klein16_LISASMBH}. 

We find that (for our specific binary evolution model) about 25 binaries will be detectable with a  signal-to-noise ratio above five after 4 years of observations. In our method, binary properties are drawn probabilistically (in Monte Carlo fashion) for each star particle according to its age and metallicity. We also vary the localization of the Sun along an annulus 8 kpc away from the Galactic center. The global properties of the binaries are identical for different realizations of the model and different positions of the Sun. We find similar results for {\bf m12b}.  {\bf m12c} predicts roughly twice as many detectable BBHs because it is much more compact, decreasing the typical distance of binaries in the main galaxy.  We caution that {\bf m12c} is more compact than the MW, however \citep{Porcel_98_MWscaleradius,SGK:2017_FIREmorph}.
The lowest frequency systems correspond to frequencies probed by pulsar timing arrays, but the corresponding strains are almost 10 orders of magnitude below the current detection limit and will remain completely undetectable for the foreseeable future with the current methods.

The population represented in Fig.~\ref{fig:f_vs_strain} results from a convolution between the star formation history and spatial structure of the galaxy and massive binary evolution. As shown in Figs.~\ref{fig:fini_Mc} and \ref{fig:hist_BH_props}, most of the binaries result from progenitor stars with $Z\simeq 0.2\Zsun$ and had an initial frequency $f_{GW}<10^{-5}$ (orbit longer than a day) and quasi-circular orbits. The high frequency binaries as well as most of the low-metallicity binaries shown in Fig.~\ref{fig:fini_Mc} have merged by now. The highest frequency systems ($f_{\mathrm{GW}} >10^{-4}$) typically have a total mass below 25 $\msun$ and come from progenitors with metallicity $Z\simeq 0.3\Zsun$.

The detection of stellar BHs without stellar companions, whether they are single BHs or BBHs, is strongly limited as such systems do not emit any electromagnetic radiation. The current strongest indications for the presence of BHs without stellar companions in the MW come from micro-lensing events with inferred lens masses up to $\simeq 10\msun$ and no detectable electromagnetic counterpart~\citep{Wyrzykowski16_OGLE}. Massive lenses lead to months-to-year long lensing events. Distinguishing between binaries with a short orbital period or single sources would be impossible making the merged systems, unmerged systems and black holes formed through any other channel (see Fig.~\ref{fig:evol}) indistinguishable. Without additional information on the distance to the lensed source, it is impossible to break the degeneracy between the mass of the lens and its distance ~\citep{Agol02_microlensing}. As such, BH candidates can only be identified in a probabilistic sense. Surveys targeting dense regions like the bulge of the galaxy are the best suited for this type of analysis ~\citep{Wyrzykowski16_OGLE}. When the distance to the source and lens are known through microlensing parallax, the mass of the lens can be determined. Surveys of the Magellanic Clouds or other nearby galaxies are the best suited for this type of search \citep{Wyrzykowski10_OGLE_MACHOS,2017arXiv171110898M}, although those events would be very rare due to the small density of the lenses. If our conclusions hold for the global distribution of BH in the MW (see Fig.~\ref{fig:evol} for all the BHs not considered here), a long-term survey of the Magellanic Clouds or M31 may be the best option to find single BHs.

\begin{figure}
\centering
\includegraphics[width = .45\textwidth]{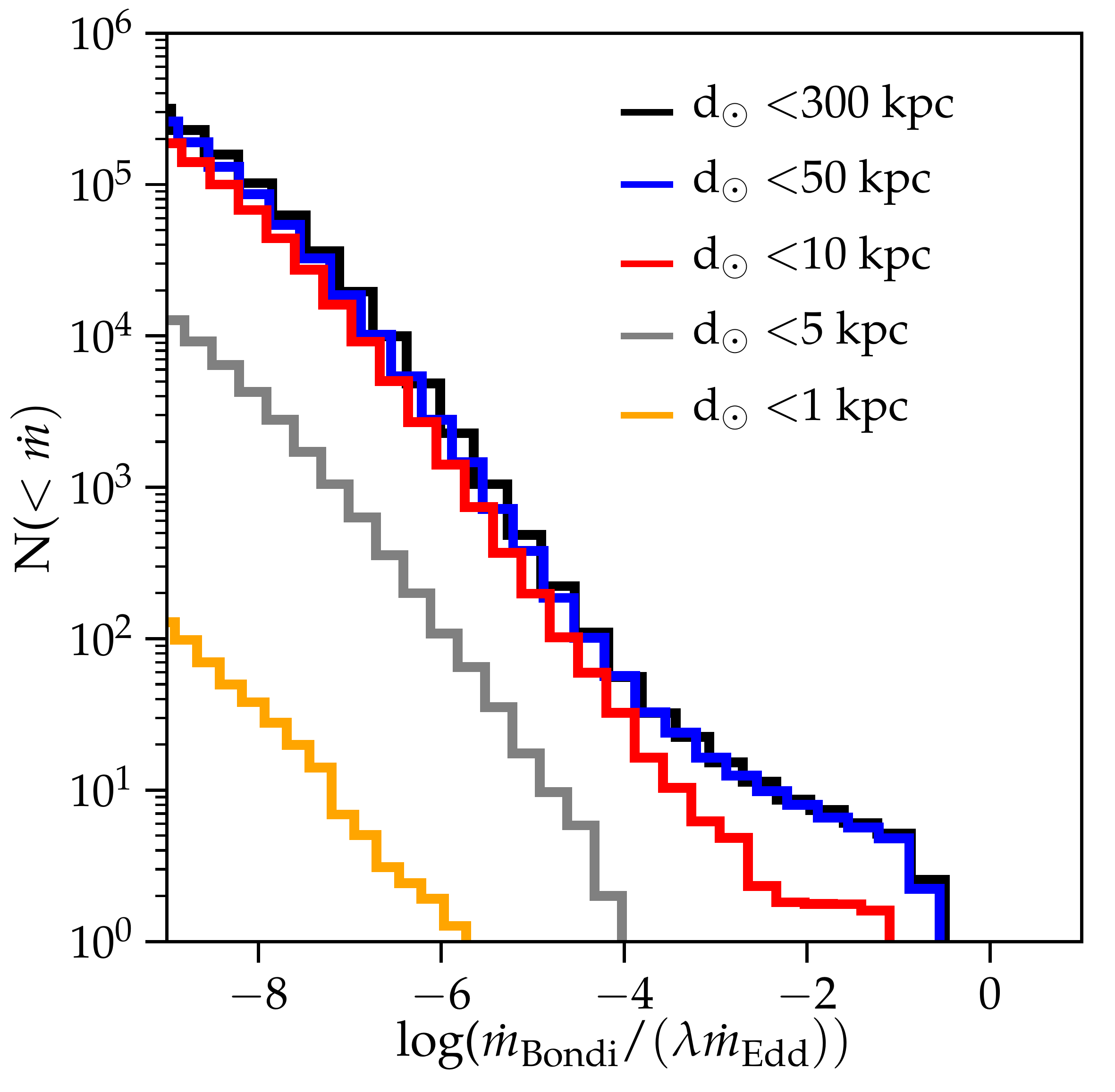}
\caption{Cumulative distribution function of the Bondi-Hoyle accretion rate of the ambient gas by merged and unmerged BBHs, normalized to the Eddington accretion rate and the Bondi-Hoyle accretion efficiency $\lambda$. Different colors indicate different distance cuts. }
\label{fig:accretion} 
\end{figure}

Faint X-ray and radio emission may be detected if BHs are accreting surrounding gas \citep{Maccarone05_radioBH}. Fig.~\ref{fig:accretion} provides the Bondi-Hoyle accretion rate $\dot{m}_{\mathrm{Bondi}}$  , with respect to the Eddington accretion rate  for all our BHs.  The Bondi-Hoyle accretion rate is given by
\begin{equation}\label{eq:bondi_hoyle}
\dot{m_{\mathrm{Bondi}}}=\lambda 4\pi(GM)^2\rho(v^2+c_s^2)^{-3/2},
\end{equation}

where $v$ is the velocity of the BH with respect to the ambient gas and $c_s$ and $\rho$ are the sound speed and density of the accreted gas.  The parameter $\lambda$ is an ``efficiency'' factor which represents	 our ignorance on detailed accretion physics.  Observational estimates suggest $\lambda=10^{-2}-10^{-3}$ based on accretion onto neutron stars \citep{Perna03_accretion}. Based on these values of the accretion efficiency,  most BH systems are expected to  accrete at less than $10^{-6}$ times the Eddington accretion rate. This is because only a small fraction of our BHs are located in the disk and bulge, close enough to dense molecular clouds, where accretion would be large. Moreover, the BHs we consider stem from old stars, which have high proper velocities with respect to the gas. Isolated black holes kicked out of binary systems are likely to have a similarly high velocity, and small accretion rate.  The systems with the highest accretions rates are located in a star-forming satellite galaxy.  The radiative properties of the accreting black holes dependent on the geometry, cooling properties, and radiative mechanisms of the accretion flow, which are highly uncertain at these low accretion rates, especially in binary BHs.  But if $\lambda$ is high in these systems, as many about 10 merged or unmerged BBH systems, mostly moving through dense molecular clouds in the inner disk or star forming satellites, could be accreting at sufficiently large rates to be detectable with all-sky X-ray surveys. Around the Milky Way,  the Magellanic Clouds are the only known satellites still actively forming stars.

\section{Discussion}\label{sec:discussion}

\subsection{An improved MW model}
The unique combination of a high-resolution cosmological simulation of a MW-mass galaxy and a binary population synthesis model allows us to predict the number of unmerged BBHs and merged BBHs in the MW, as well as their mass distribution and localization. From there, we can determine possible detections, with electromagnetic telescopes or gravitational wave detectors.  

Our MW model is based on a cosmological simulation of a MW-mass galaxy down to $z=0$ \citep{Wetzel2016}, with a mass resolution of $\simeq 7000 \msun$. We find that the number, mass distribution and spatial distribution of the merged BBHs are nearly identical in a simulation of the same galaxy with a factor of 8 fewer particles (mass resolution of $\simeq56000\msun$). In contrast, the unmerged BBHs are 15 per cent more numerous in the lower resolution simulation, due to somewhat higher star formation after $z\simeq 1.5$. This results into more systems from progenitors with $Z>0.1 \Zsun$, and a slightly lower mean mass for the BBH systems. The spatial distribution of the BHs is the same in both cases, aside from the satellites, which produce fewer BH in the lower-resolution simulation. Altogether, this suggests our results are not strongly sensitive to the resolution of the simulation.

\begin{figure*}
\centering
\includegraphics[width = .95\textwidth]{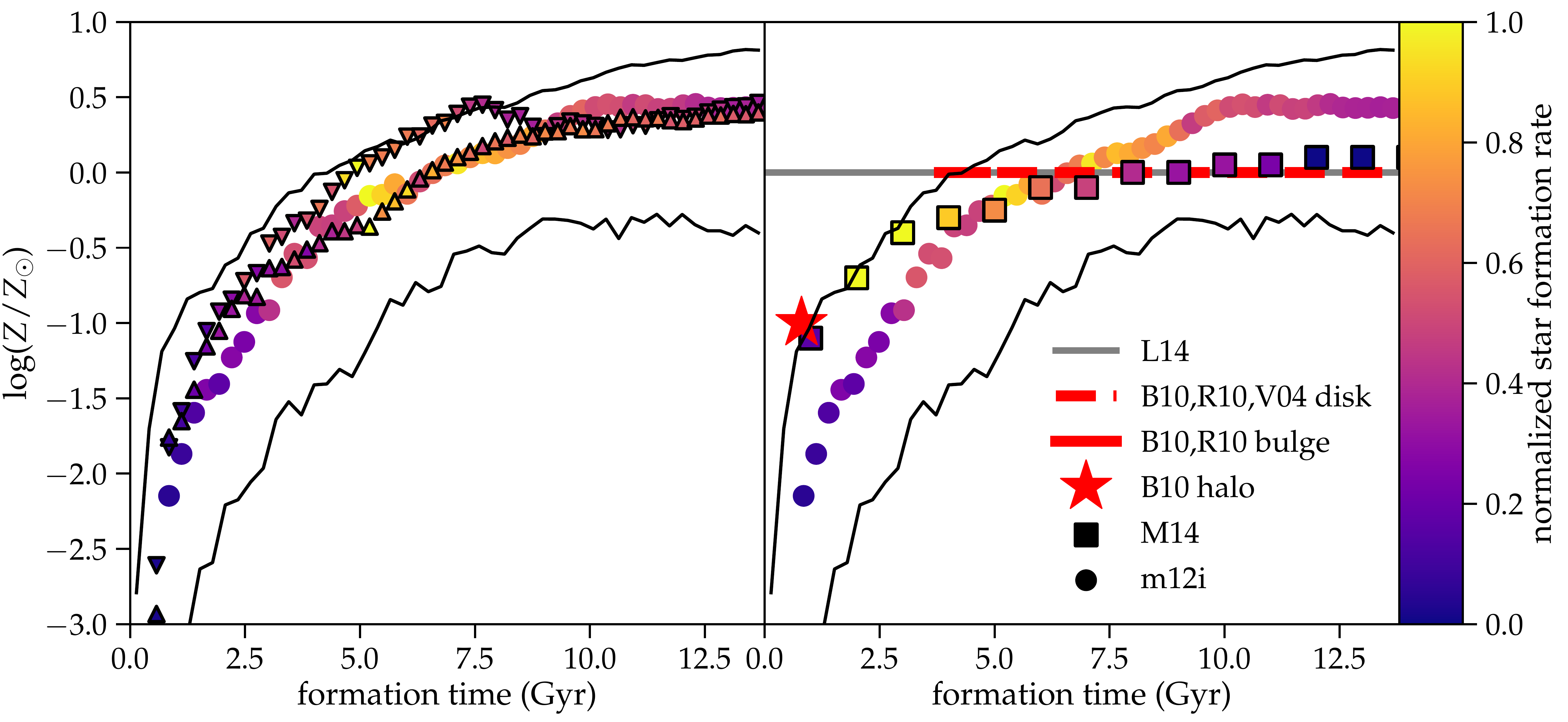}
\caption{Comparison of the star formation rate as a function of metallicity and time used in the \textbf{m12i} simulation compared to  other simulations of MW-type galaxies (left) from the FIRE project and previous models for the MW used to derive BH populations in the Galaxy (right). In both plots, the  circles show the median values in the \textbf{m12i} simulation, color-coded  by their relative star formation rate ($\mathrm{SFR}(t,Z)/\max\mathrm{SFR}$) and the black lines show the 95$\%$ scatter in metallicity for stars forming at the same time. On the left plot, monochromatic lines show models that assume a constant star formation rate (e.g. \citet{Voss03_BPS,Ruiter2010_WD,Belczynski2010_LISA,Liu14_LISA}), while colored points are shaded by the relative star formation rate at a given time (e.g. \citealt{Mennekens2014_DCO}). Burst of star formation are shown with stars symbols.  On the right plot, we show the \textbf{m12b} simulation with inverted triangles and \textbf{m12c} simulation with triangles. }
\label{fig:Latte_vs_others} 
\end{figure*}

More importantly, the simulation provides a fully cosmological model for the formation of a Milky Way-mass galaxy, self-consistently modeling its stellar disk, satellite population, and stellar halo, each with cosmologically driven formation histories and self-consistent metal enrichment. This is a significant improvement over previous estimates of the population in the MW. The left panel of Fig.~\ref{fig:Latte_vs_others} compares the star formation rate as a function of time and metallicity in the simulation, versus other models that are used to compute the binary population in the MW. Most models assume a constant star-formation rate at Solar metallicity (straight monochromatic lines in Fig.~\ref{fig:Latte_vs_others}) in the disk and bulge, if the latter is included at all. When present, the halo is assumed to come from a single star burst at $Z=0.1\Zsun$ \citep{Voss03_BPS,Belczynski2010_LISA,Ruiter2010_WD,Liu14_LISA}. All these models neglect super-solar metallicity star formation. Although this does not impact BBH formation, it may impact the formation of binary neutron stars and/or white dwarfs \citep{Nelemans01_WD,Ruiter2010_WD}. 
Only \citet{Mennekens2014_DCO} include time-variable, metallicity-dependent star formation (colored squares), although their model globally over-predicts early star formation and underestimates the variation in the mean metallicity over time compared to observational constraints in MW-mass galaxies (e.g. \citet{Behroozi13_SFRz}. None of the models in the literature include scatter in the metallicity at a given time, while the dashed lines (and also Fig.~\ref{fig:SFR_Z_m12}) and observations \citep{GarciaPerez17_APOGEE} show that there is a typical range of about a dex between the highest and lowest metallicity at a given time. This is crucial as the lowest metallicity systems most significantly contribute to the formation of BBHs.

Models in the the literature also neglect the crucial impact of galactic mergers. In Fig.~\ref{fig:hist_BH_props} we have shown that about a third of the BBHs present in the MW come from progenitors formed in a satellite galaxy. At the highest masses, more than 60 per cent of the BBHs comes from ex situ formation.   This is somewhat different from the findings of \citet{Chakrabarti17_HIdisk} who predict that most of the progenitors of BBH mergers come from the outer disk of massive galaxies. Neglecting galactic mergers underestimates the global BBH population, specifically high mass binaries and binaries in the galactic halo.

\citet{Belczynski2010_LISA}  assume stars form at solar metallicity over the past 10 Gyrs with a small contribution from the bulge and halo (at $Z=0.1\Zsun$). In contrast, the simulation shows that the majority of stars formed during the last 7 Gyrs have a super-solar metallicity, and are unavailable to BBH formation. On the other hand, they neglect most of the star formation between $z=1.5-3$, where we predict the majority of the BBHs originate (see Fig.~\ref{fig:SFR_Z_m12}).  This may explain why their model B (which is similar to our BPS model) predicts about half as many binaries as our model even though the assumptions on the binary evolution are very similar.  They also neglect the contribution from ex-situ stars, effectively missing 40 per cent of the binaries. We also predict more \textit{LISA} detections because, on average, our binaries stem from more metal-poor stars and produce more massive black holes.

We also show the star formation history of two other simulations of MW-mass galaxies run with the same resolution and physics (right panel of Fig.~\ref{fig:Latte_vs_others}). The \textbf{m12b} simulation has a 27 per cent higher present-day stellar mass than \textbf{m12i}, but produces fewer BHs. This is because most of its star formation is at too high metallicity to significantly produce BBHs (see inverted triangles in Fig.~\ref{fig:Latte_vs_others}). \textbf{m12c}, meanwhile, has 7 per cent less present-day stellar mass, but also produces 15 (13) per cent fewer merged (unmerged) BBHs. This is because its low metallicity star formation around $z\simeq2$ is lower than in our reference simulation. The global properties (mass, progenitor metallicity and spatial distribution) of the BHs are similar in the three simulations. The comparison with different MW-mass galaxies highlights the robustness of our calculation, and indicates that our simulation is a reliable representation of the Milky Way galaxy, even though it is not an exact reproduction. The comparison also shows that although the present-day stellar mass provides a first indication of the BBH content of a galaxy, the details of its star formation history, and more specifically its metallicity can have an important effect.  

Our host galaxy has a higher present-day star formation rate than the MW. As recent star formation has a limited contribution to BBH formation (see Fig.~\ref{fig:SFR_Z_m12}), we do not expect this to influence the quality of our Milky Way model. Although the global properties of the satellites are consistent with observations \citep{Wetzel2016}, the simulation does not show a massive satellite like the Large Magellanic Cloud. The latter still presents star formation with $Z\simeq 0.25\Zsun$ \citep{harris09_SFR_LMC} and is thus a prime candidate for BBH formation. As such, our estimate of the number of UBBH and MBBH within 300~kpc is likely a lower limit.

\subsection{Pathways towards detections}
The accurate spatial model in our fully cosmological simulation, with respect to simplified disk and halo models, allows us to provide some predictions of the detectability of the sources with electromagnetic or gravitational signatures.  Although BH natal kicks are included in our binary evolution model, we do not model the impact of BH kicks on the location and proper motion of the BHs in the galaxy simulation. Black holes kicks are assumed to be smaller than neutron star kicks, due to material falling back. As we are specifically focusing on BBHs surviving SN kicks, we are biased toward the BBHs with small natal kicks (otherwise the binary would have been disrupted). Additionally, most of our systems are found in the stellar halo, which is dominated by random motions, where BH natal kicks are likely to have a limited impact of the statistical properties of the total BBH distribution. Systems formed in the disk may be kicked into the halo, especially if the kicks occur before most the mass of the host galaxy is accreted ($z\lesssim 3$). For systems formed in satellite galaxies, kicks may be powerful enough to escape the shallow potential well of the dwarf galaxy but it is unlikely that they would be able to escape the potential of the main host. As a result, the distribution of ex-situ systems may be more randomized, and BBHs may not closely trace stellar streams and satellites. Overall, we expect most of the merged and unmerged BBHs to be in the galactic halo, and we argue that this result is robust to our assumption that BBHs trace the location of their parent star particles.

The prospects for detecting black holes without stellar companions remain limited. Still, with a refined model of the star formation history in the MW, non-detections with X-ray, radio and microlensing surveys will put constraints on binary evolution models. Conversely,  a much higher number of detected systems will be equally constraining, and may inform us of additional BBH formation channels \citep{Marchant:2016,Rodriguez:2015}.   Our understanding of massive binary evolution is still very uncertain, especially the modeling of stellar mass loss through different evolutionary phases, the properties of common-envelope mass transfer, and supernova kicks. Specifically, we have assumed that common envelope interactions during the Hertzprung gap result in stellar mergers, which may underestimate the BBH production rate. Conversely, we assume the BH natal kicks are reduced with respect to neutron star natal kicks, which may overestimate the number of systems surviving both supernova explosions. As our work highlights the importance of improved galactic models, we do not explore the wide parameter space of current binary models. The different BBH merger rates predicted by \citet{Mapelli17_Illustris} show how different binary evolution models will eventually be ruled in or out by observational data, provided appropriate models are used for the star formation. 

Based on our assumptions for binary evolution, we predict that \textit{LISA} will detect about 25 BBH within the MW halo. The frequency of these systems will not vary over the duration of the \textit{LISA} mission, which means that the measured chirp mass will be degenerate with the distance of the system (see Eq.~\ref{eq:strain}) and discerning it from a closer neutron star binary may not be possible. While certain aspects of binary evolution are likely to be revised, we emphasize that the BBH predictions of our model are somewhat on the high end (see a discussion in \citealp{Lamberts16_GW150914}) and that fine-tunning it in order to produce more galactic BBHs may overproduce the observed BH merger rate \citep{LIGO_GW170608}.  Unless a mechanism allows for the formation of BBHs at close-to solar metallicity, BBHs are strongly biased towards the galactic halo, which strongly reduces their GW signal on Earth relative to disk populations.  Even a single detection with accurate localization will provide information on the binary evolution mechanism and the conditions of its formation. 

We find that less than a million binary black holes have merged in the MW by the present day. Focusing on the mergers within the last 2 Gyrs, we find a merger rate of $\simeq 10^{-5}$ yr$^{-1}$. Our model, which assumes a binary fraction of unity, is on the higher end of the measured rate.  We find the mean mass of the systems to be 29 $\msun$, which is in line with the announced detections \citep{LIGO_O1_BBH,LIGO_GW170104,LIGO_GW170814,GW170608}.  We find that sources with $M_{\mathrm{tot}}>50\msun$ represent 8 per cent of the merged systems, meaning there could be roughly 40,000 such BBHs in our Galaxy. The progenitors of these systems likely formed in a satellite galaxy and are now present in the halo. Similarly to \citet{Lamberts16_GW150914}, we find that mergers in MW-mass galaxies primarily stem from progenitors with $Z\simeq 0.1\Zsun$.

Initial studies predicted roughly 10$^8$ BH in the Milky Way (single and binary), based on stellar evolution models \citep{Shapiro_Teukolsky,VandenHeuvel92_BHinMW} and yields from massive stars \citep{Samland98_yields}. Based on metallicity-dependent star formation models of galaxies of all masses, \citet{Elbert17_BHcount} predict $\simeq 10^8$ black holes in the MW, with 10 per cent of them above 30 $\msun$. Our model predicts about a million binary black holes, and less than a million merged systems. Our calculation only accounts for binary black holes from field binaries, and we only track systems where both stars form a BH, where the binary does not undergo a stellar merger, and where the binary survives the natal kicks. As such, we do not account for black holes with lower mass companions (white dwarfs, neutron stars, or low-mass stars). Thousands of black holes with stellar companions will likely by detected by the astrometric \textit{GAIA} satellite \citep{Mashian17_BHGAIA,Breivik17_BHGAIA} and possibly their H$\alpha$ emission \citep{Casares17_BH_Halpha}. Given that even at the lowest metallicity, only 8 per cent of the massive binaries turn unto a BH binary, there could be 10 times more black holes that have been kicked out of a binary. As such, the first electromagnetic detection of isolated black holes in the MW will most likely be a single black hole. A more accurate determination of their masses, proper motions, spatial distribution and ultimately, detectability, is left for a further study.

\section{Conclusions}\label{sec:conclusion}

In this paper we provide the first estimates of the isolated black hole population resulting from binary black hole systems. It is based on the combination of a high resolution cosmological simulation of a MW-mass galaxy and a binary population synthesis model. The simulation provides a physically-motivated, metallicity-dependent star formation history as well as a complete description of the galactic morphology and merger activity over time. The simulation models both resolved and unresolved turbulent metal diffusion, which provides a realistic metal \textit{distribution}, rather than just the average value \citep{Ma17_metalMW,Escala2017_dmdf}. The stellar metallicity is a key parameter for massive binary evolution. Using a standard binary evolution model, we compute a metallicity-dependent library of binary black holes for 13 metallicities between $Z=0.0005 \Zsun$ and $Z=1.6\Zsun$, then match them to the stars in the simulation. This provides a self-consistent distribution of black holes within 300 kpc of the center of the galaxy, including their localization, masses, orbital properties and the properties of their stellar progenitors.  The main properties of these binaries are summarized below:

\begin{itemize}
\item We find that $6.8\times10^5$  binary black hole binaries have already merged in our MW model and $1.2\times 10^6$ systems are still in binary black holes. Our Milky-Way like galaxy has turned 0.07 per cent of its $z=0$ stellar mass into binary black holes, including the ones that have merged already. 
\item The mean progenitor metallicity of the merged (unmerged) systems is  $Z=0.13$ $(0.25)$~$\Zsun$, and only 1 per cent of the binaries come from super-solar metallicity progenitors. This means that most of the stellar mass in MW-mass galaxies is effectively unavailable for the formation of black hole binaries. The binary systems thus strongly trace star formation around $z\simeq 2$ while the already-merged systems mostly trace star formation from $z>2.$
\item The strong dependence on low metallicity star formation results in half of the binaries (merged or not) being located beyond 10 kpc of the galactic center. In comparison, 90 per cent of the stellar mass is located within 10 kpc. The galactic halo, streams and satellite galaxies are rich in BBHs.
\item We find about 40,000 merged binaries with masses comparable to the mergers detected with the  first LIGO detection. Consistent with \citet{Lamberts16_GW150914}, we find that  these systems were typically formed outside of the MW and are now in the halo or still in their host satellite galaxy. They stem from progenitors with metallicity below $0.1\Zsun$. 
\item The detection of merged and unmerged binary black holes without stellar companions will remain a challenge for the foreseeable future.  Our binary evolution model predicts that 25 binaries could be detected with \textit{LISA}, but that most of the binaries have an orbit that is too wide and/or are too distant for \textit{LISA} to detect their gravitational wave emission. Depending on the radiation efficiency and accretion mechanism,  the accretion of surrounding gas could lead to detections with  all-sky radio or X-ray surveys like \textit{SKA} or \textit{eROSITA}. A few systems may be detected by microlensing surveys. In any case, observational constraints will lead to a better understanding of massive binary evolution, which is still poorly understood.
\item About a third of the binary black holes were not initially formed in the galaxy, and have been brought in by the accretion of satellite dwarf galaxies. 60 per cent of the systems with total mass above 60 $\msun$ were formed ex situ. This highlights the importance of accounting for galactic mergers when predicting black hole populations and merger rates in MW-mass galaxies.
\item The mean total mass is 29 $\msun$ for the merged systems and a total mean mass of 23 $\msun$ for the binary systems. Roughly 10 per cent of the systems have total masses above 50 $\msun$ and about one per cent of the systems have a mass above 80 $\msun$.
\item We provide online data including a table of the star formation rate as a function of metallicity and time. The latter can be used to combine the galaxy model with different binary evolution models. We also provide the properties of the binary black holes from our model, which can be used to derive observational signatures from lensing and/or accretion.
\end{itemize}

This paper provides the first estimate of the population of binary black holes and their merger remnants in a MW-mass galaxy that incorporates a realistic star-formation history  and galactic halo structure and merger history based on  a hydrodynamic simulation. This allows us to reliably estimate low metallicity star formation and localize binary black holes. Based on this precise galactic model, future detections of such systems in our vicinity will then allow us to constrain key parameters of massive stellar binary evolution.

\section*{Acknowledgements}

Astrid Lamberts would like to thank V. Ravi, H. Vedantham, M. Heida, C. Henderson, Y. Shvarzvald, S. Novati and S. Taylor for discussions about observational implications of this work and D. Clausen for his help with the BPS models.  Numerical calculations were run on the Caltech compute cluster ``Wheeler,'' allocations from XSEDE TG-AST130039 and PRAC NSF.1713353 supported by the NSF, and NASA HEC SMD-16-7592. 

Support for AL and PFH was provided by an Alfred P. Sloan Research Fellowship, NASA ATP Grant NNX14AH35G, and NSF Collaborative Research Grant 1715847 and CAREER grant 1455342.  
Support for SGK was provided by NASA through Einstein Postdoctoral Fellowship grant number PF5-160136 awarded by the Chandra X-ray Center, which is operated by the Smithsonian Astrophysical Observatory for NASA under contract NAS8-03060.  EQ was supported in part by NSF grant AST-1715070 and a Simons Investigator Award from the Simons Foundation.  JSB was supported by NSF grant AST-1518291 and by NASA through HST theory grants (programs AR-13921, AR-13888, and AR-14282.001) awarded by STScI, which is op- erated by the Association of Universities for Research in Astron- omy (AURA), Inc., under NASA contract NAS5-26555. CAFG was supported by NSF through grants AST-1412836, AST-1517491, AST-1715216 and CAREER award AST-1652522 and by NASA trough grant NXX-15AB22G. AW was supported by NASA through grants HST-GO-14734 and HST-AR-15057 from STScI.
DK acknowledges support from NSF grants AST-1412153 and AST-1715101 and the Cottrell Scholar Award from the Research Corporation for Science Advancement. 
RES was supported by an NSF Astronomy \& Astrophysics Postdoctoral Fellowship under grant AST-1400989. This study was initiated during K. Drango's ``Freshman Summer Research Internship'', organized by the Caltech Center for Diversity.




\bibliographystyle{mnras}

\bibliography{sources}







\bsp	
\label{lastpage}
\end{document}